\documentclass{mn2e}
\usepackage{color}
\usepackage{graphicx}
\usepackage{dcolumn}
\usepackage{bm}

\begin{document}

\title[Inclination- and dust-corrected galaxy parameters]
{Inclination- and dust-corrected galaxy parameters: 
Bulge-to-disc ratios and size-luminosity relations}

\author[Graham \& Worley]
{Alister W.\ Graham$^{1,2}$ and C.\ Clare Worley$^{3}$\\
$^1$Centre for Astrophysics and Supercomputing, Swinburne University
of Technology, Hawthorn, Victoria 3122, Australia.\\
$^2$Corresponding Author: AGraham@astro.swin.edu.au\\
$^3$University of Canterbury, Department of Physics and Astronomy, Private Bag 4800, Christchurch 8020, New Zealand.
}

\date{Received 2008 Jan 28; Accepted 2008 May 23}

\input{psfig}
\maketitle
\label{firstpage}

\begin{abstract}

While galactic bulges may contain no significant dust of their own, the dust
within galaxy discs can strongly attenuate the light from their embedded
bulges.  Furthermore, such dust inhibits the ability of
observationally-determined inclination corrections to recover intrinsic
(i.e.\ dust free) galaxy parameters.  Using the sophisticated 3D radiative
transfer model of Popescu et al.\ and Tuffs et al., together with Driver et
al.'s recent determination of the average face-on opacity in nearby disc
galaxies, we provide simple equations to correct (observed) disc central
surface brightnesses and scalelengths for the effects of both inclination
and dust in the $B, V, I, J$ and $K$ passband.  We then collate and
homogenise various literature data sets and determine the typical intrinsic
scalelengths, central surface brightnesses and magnitudes of galaxy discs as
a function of morphological type.
All galaxies 
have been carefully modelled in their respective papers with a S\'ersic
$R^{1/n}$ bulge plus an exponential disc.  Using the bulge magnitude corrections
from Driver et al., we additionally derive the average, dust-corrected,
bulge-to-disc flux ratio as a function of galaxy type.  With values typically
less than 1/3, this places somewhat uncomfortable constraints on some current 
semi-analytic simulations.  Typical bulge sizes, profile shapes, surface
brightnesses and deprojected densities are provided. 
Finally, given the two-component nature of disc galaxies, we present
luminosity-size and (surface brightness)-size diagrams for discs and bulges. 
We also show that the distribution of elliptical galaxies in the
luminosity-size diagram is not linear but strongly curved.

\end{abstract}



\section{Introduction}

Although the bulge-to-total ($B/T$), or bulge-to-disc ($B/D$), flux ratio was
the third, not prime, galaxy-morphology criteria employed by Sandage (1961),
its systematic behaviour along the Hubble sequence is well known 
(e.g.\ Boroson 1981; Kent 1985; Kodaira, Watanabe \& Okamura 1986; Simien \&
de Vaucouleurs 1986).  While one-third (two-thirds) of the stellar mass
density in the Universe today is known to reside in bulges (discs) (e.g.\
Driver et al.\ 2007), the actual allocation of stars as a function of galaxy
type is not securely known.  This is however a quantity of interest.  If
(large) bulges, including elliptical galaxies, are the result of mergers,
while discs formed via the gravitational collapse of a rotating proto-galactic
cloud, or from the accretion of gas around a pre-existing galaxy, then the
$B/D$ ratio reflects the dominance of a galaxy's formation mechanisms 
(e.g.\ Navarro \& Benz 1991; Steinmetz \& Navarro 2002 and references therein).
The reliable separation of bulge and disc stars is also important because of
the connection between supermassive black hole mass and the physical
properties of the host bulge (see the review in Ferrarese \& Ford 2005).
%

For many years it was believed that the bulges of disc galaxies were
universally described by de Vaucouleurs empirical $R^{1/4}$ model (de
Vaucouleurs 1948, 1958, 1978).  While deviations were occasionally noted for
individual galaxies, such as the Milky Way (Kent, Dame \& Fazio 1991), the
above belief only began to change in earnest after Andredakis \& Sanders
(1994) showed that the exponential model provided a better description to the
distribution of stellar light in the bulges of 34 late-type spiral galaxies; a
result reaffirmed by de Jong (1996a) and Courteau, de Jong \& Broeils (1996)
using larger samples.
%
%
However the fuller renaissance, if one can use such a term, started with
Andredakis, Peletier \& Balcells (1995) who showed that S\'ersic's (1963,
1968) $R^{1/n}$ model was particularly well suited to
describing the light-profiles of bulges.  This result was subsequently
confirmed by other investigations (e.g.\ Iodice, D'Onofrio \& Capaccioli 1997,
1999; Seigar \& James 1998; Khosroshahi, Wadadekar \& Kembhavi 2000) and
S\'ersic's model has since become the standard approach for describing bulges
in both lenticular and spiral galaxies (e.g.\ D'Onofrio 2001; Graham 2001a;
M\"ollenhoff \& Heidt 2001; Prieto et al.\ 2001; Castro-Rodr{\'{\i}}guez \&
Garz\'on 2003; MacArthur et al.\ 2003; Balcells et al.\ 2007; Carollo et al.\
2007).

\begin{figure}
\includegraphics[angle=270,scale=0.37]{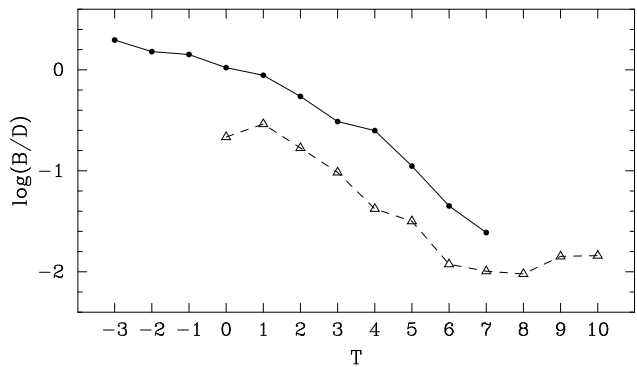}
\caption{
Average logarithm of the $B$-band, bulge-to-disc flux ratio as a function of 
galaxy morphological T-type.  The solid line traces data from Simien \& de Vaucouleurs (1986)
who used an $R^{1/4}$ bulge $+$ exponential disc model.   
The $R^{1/4}$ model is known to over-estimate the bulge flux
when the bulge is better described with a S\'ersic $R^{1/n}$ profile having
$n<4$ (e.g.\ Trujillo, Graham \& Caon 2001). 
The dashed line traces data from Graham (2001a) who applied an 
$R^{1/n}$ bulge $+$ exponential disc model to de Jong \& van der Kruit's
(1994) roughly face-on ($b/a > 0.625$) sample of galaxies. 
Both curves, however, still need to be properly adjusted for dust
attenuation. 
}
\label{Fig_old}
\end{figure}

Today, bulges are observed to follow a roughly linear magnitude-S\'ersic index
relation, with most bulges having S\'ersic indices $n < 4$ (e.g.\ Graham
2001a; MacArthur et al.\ 2003).  Indeed, Balcells et al.\ (2003, see their
figure~1a) revealed that even lenticular galaxies typically have values of $n$
around 2, a result which has subsequently been echoed by others (e.g.\
Laurikainen et al.\ 2006).  
Application of the average $B/D$ ratios derived from studies which used
$R^{1/4}$ bulge $+$ exponential disc decomposition introduces a bias into the
separation of bulge and disc flux when a bulge does not have an $R^{1/4}$
profile (e.g.\ de Jong 1996b; Trujillo et al.\ 2001; Brown et al.\ 2003).
Specifically, given that the bulges of most disc galaxies have S\'ersic
indices $n<4$, application of the $R^{1/4}$ model (with its higher central
concentration of stars and greater tail at large radii) assigns too much of
the galaxy flux to the bulge.  Figure~\ref{Fig_old} shows the average $B$-band
$B/D$ ratios, as a function of galaxy type, from the data in Simien \& de
Vaucouleurs (1986), whose bulge magnitudes were obtained using an $R^{1/4}$
model.
Also shown in Figure~\ref{Fig_old} are the $B$-band $B/D$ ratios obtained by 
Graham (2001a) using an $R^{1/n}$ bulge model.  
Figure~\ref{Fig_old} is not however the end of the story, as neither study
addressed the issue of attenuation of the bulge flux due to dust in the disc
of the galaxies. 

This brings us to our second, and often over-looked, concern. 
While bulges may contain little to no dust of their own --- and have therefore
in the past not had their flux corrected for dust --- the galaxy discs which
effectively cut them in half do contain dust.  Consequently, particularly at
optical wavelengths, one actually sees very little of the star light from the
portion of a bulge on the far side of a disc.  The extent to which the
bulge and disc flux is dimmed depends on both the observed wavelength and the
inclination of the (dusty) stellar disc.  The
reduction to the observed magnitude of the bulge and disc components can be as
high as 2-3 mag in the $B$-band (e.g.\ Driver et al.\ 2007).  In spite of
this, the overwhelming majority of published (optical) bulge magnitudes, and
thus bulge-to-disc flux measurements, have not taken this into account and are
therefore in considerable error.

On the other hand, for over the last decade most studies have corrected, at
least in part, for the influence of dust on the discs.  This has been
accomplished by noting how parameters such as magnitude and surface brightness
change with viewing angle and adjusting the observed values to those which
would be observed with a face-on orientation (e.g.\ Valentijn 1990; de
Vaucouleurs' et al.\ 1991; Giovanelli et al.\ 1995; Tully et al.\ 1998; Graham
2001b; Masters et al.\ 2003).  To obtain the intrinsic (dust-free) values of
course requires an additional step: namely, correcting the face-on galaxy
parameters for the influence of dust.

Sophisticated 3D dust/star galaxy models, which 
incorporate both clumpiness and explicit treatment of various grain
compositions and sizes, now exist. 
The model of Popescu et al.\ (2000) and Tuffs et al.\ (2004) self-consistently
explains the UV/optical/FIR/sub-mm emission from galaxies.  Indeed,
application of their dust model to 10,095 galaxies from the Millennium Galaxy
Catalog (e.g. Allen et al.\ 2006) perfectly balances the amount of star light
absorbed by dust in the Universe today with the amount re-radiated at infrared
and sub-mm wavelengths (Driver et al.\ 2008).  For almost a decade these
models accounted for the fact that the dust is non-uniformly distributed,
allowing for high extinction in the central regions, intermediate extinction
in the spiral arms, and semi-transparent interarm regions which accounted for
the detection of background galaxies.  Their detailed model not only allows
one to accurately correct for inclination-dependent extinction, but also for
the additional extinction which is present when viewing galaxies with face-on
orientations.

The one free parameter in their dust model which provides the calibration 
is the central, face-on, $B$-band opacity $\tau_B^f$. 
Fitting a range of simulated galaxies with varying integer-values of 
$\tau_B^f$, M\"ollenhoff, Popescu \& Tuffs (2006)
provided tables to correct observed disc parameters to their face-on,
dust-free values.  Combining their tables with the statistically determined
average opacity $\tau_B^f = 3.8 \pm 0.7$ reported by Driver et al.\ (2007),
Section~\ref{Sec_dust} of this paper derives two new equations which can be
used to easily correct the observed $B, V, I, J$ and $K$-band disc
scalelengths and central surface brightnesses for both
inclination and the attenuating effect of dust, providing intrinsic, face-on,
dust-free values.  The above face-on, central $B$-band opacity from Driver et
al.\ (2007) was obtained by matching the observed inclination-attenuation
relations for the Millennium Galaxy Catalog data with the dust models of Tuffs
et al.\ (2004).
%
%


In Section~\ref{Sec_Cat} we introduce the galaxy data sets which shall be used
in our analysis of disc galaxy structural parameters, 
while Section~\ref{SecHomo} describes the methodologies
adopted to bring this data onto a uniform system.
Section~\ref{Sec_Single} provides tables of the mean intrinsic structural
parameters as a function of galaxy type in various passbands.  This
encompasses disc scalelengths, central surface brightnesses and magnitudes. 
Average $K$-band bulge effective radii, effective surface brightnesses,
S\'ersic indices and magnitudes are also listed.  In addition, and also as a
function of galaxy type, we provide the median bulge-to-disc size ratio in the
$K$-band and the median dust-corrected bulge-to-disc flux ratio in the
near-infrared and various optical bands.
 
Finally, a selection of bivariate plots are shown in Section~\ref{Sec_Double}.
In addition to (surface brightness)-size diagrams for discs and bulges, we
provide new expressions for the size-luminosity relations of discs in early-
to mid-type disc galaxies.  We also present the size-luminosity relation for
bulges and elliptical galaxies, emphasizing that it is not a linear relation.

While inclination corrections have been applied for many years, and are still
necessary today (e.g.\ Bailin \& Harris 2008; Maller et al.\ 2008; Unterborn
\& Ryden 2008), they only remove one of several systematic biases that cause
the observed flux distribution to misrepresent the true (intrinsic) stellar
distribution of galaxies.  Allowances for non-homology through the use of
S\'ersic (bulge) models and separate dust corrections to both the disc and
bulge are vital if we are to know the intrinsic physical properties of
galaxies.  It is hoped that the distributions and trends shown here, which
have been acquired after dealing with the above three issues, as done in
Driver et al.\ (2007), will provide valuable constraints for simulations
of galaxy formation which are used to aid our understanding of galaxy
evolution (e.g.\ Cole et al.\ 2006; Springel \& Hernquist 2003; Almeida et al.\ 2008; 
Croft et al.\ 2008, and references therein).

A Hubble constant of 73 km s$^{-1}$ Mpc$^{-1}$ has been used when
$H_0$-independent distances were not available.

\section{Corrections for dust and inclination} \label{Sec_dust}

\subsection{Disc scalelengths and central surface brightnesses}

The discs of disc galaxies contain dust, in particular various mixes
of graphite and silicate. 
This reduces the amount
of (ultraviolet, optical and near-infrared) light which escapes from such galaxies. 
The distribution of this dust is known not to be uniform, 
with the centres of discs containing more light-absorbing particles than their
outskirts (Boissier et al.\ 2004; Popescu et al.\ 2005).  
Such dust 
not only decreases the observed central surface brightnesses, $\mu_0$, 
of the discs, but the radial gradients in the opacity also make 
the observed disc scalelengths, $h$, greater than their intrinsic values.

\begin{figure*}
\includegraphics[angle=270,scale=0.77]{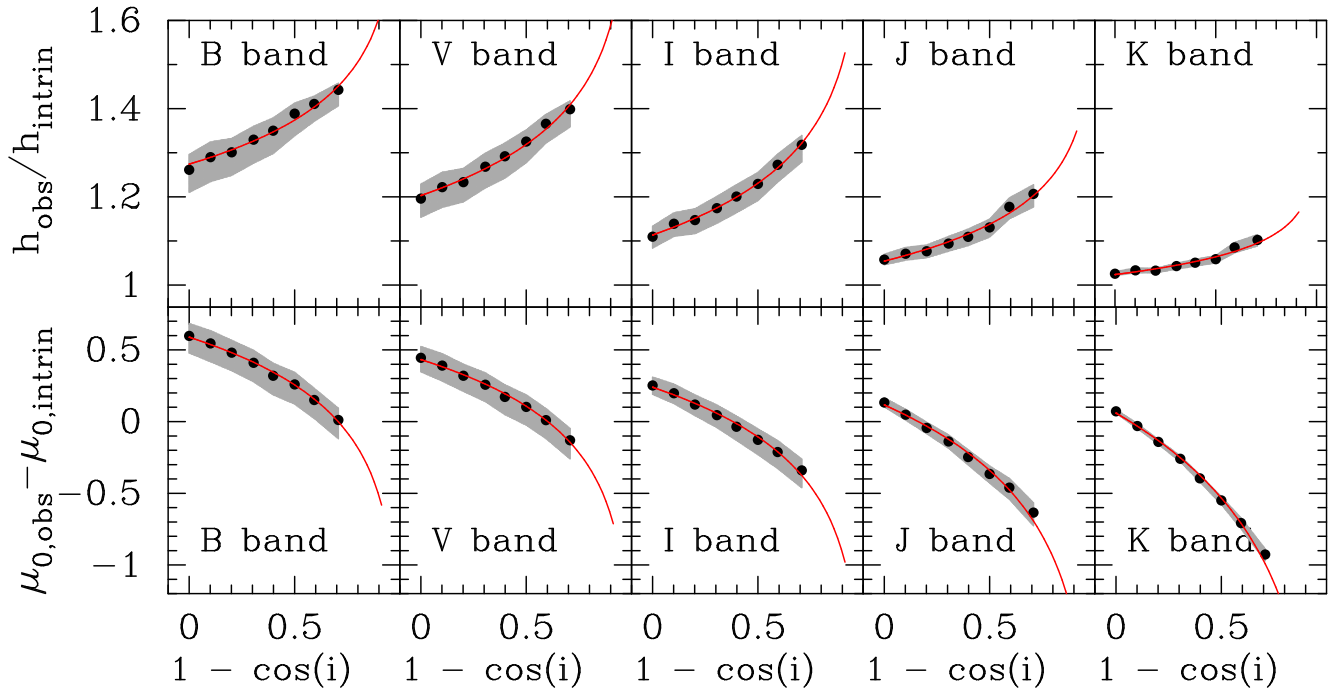}
\caption{Inclination-attenuation corrections to the scalelength ($h$) and central surface
  brightness ($\mu_0$) of simulated disc galaxies.  The data points and the associated
  shaded area have been extracted from M\"ollenhoff et al.\ (2006) using the 
  statistically determined, 
  face-on $B$-band opacity $\tau_B^f = 3.8 \pm 0.7$ (Driver et al.\ 
  2007).  The data points show the difference between the observed values 
  at some inclination, $i$, and the intrinsic face-on ($i=0$) 
  values which would be observed in the absence of dust.  The solid line is an 
  empirical fit (see equations~\ref{Eq_mu} and \ref{Eq_h}) using the code FITEXY from
  Press et al.\ (1992).  The parameters of the fit are given in
  Table~\ref{Tab_dust}.  From $0.7 < 1 - \cos i < 1.0$, the curve is an
  extrapolation of the model that was fitted to the data at smaller
  inclinations; as such it may not be applicable over this high-inclination range.
}
\label{fig-Dust}
\end{figure*}

\begin{table}
\caption{
Dust correction parameters for equations~\ref{Eq_mu} and \ref{Eq_h}, which
have themselves been fitted to the simulated data in Figure~\ref{fig-Dust}. 
These equations can be used to correct the observed scalelengths and central
surface brightnesses of discs for the effects of dust and varying inclination, 
reproducing the stellar distributions' intrinsic (i.e.\ dust-free) face-on values. 
\label{Tab_dust} 
}
\begin{tabular}{ccccc}
\hline
Filter  &  $a_{\lambda}$  &  $b_{\lambda}$  &  $c_{\lambda}$  &  $d_{\lambda}$  \\
\hline
$B$  &     $0.59\pm0.06$  &  $0.44\pm0.09$  &  $1.27\pm0.02$  &  $0.33\pm0.07$  \\
$V$  &     $0.43\pm0.05$  &  $0.43\pm0.08$  &  $1.20\pm0.02$  &  $0.39\pm0.07$  \\
$I$  &     $0.24\pm0.04$  &  $0.46\pm0.07$  &  $1.11\pm0.02$  &  $0.39\pm0.06$  \\
$J$  &     $0.12\pm0.02$  &  $0.61\pm0.05$  &  $1.05\pm0.01$  &  $0.28\pm0.04$  \\
$K$  &     $0.06\pm0.01$  &  $0.79\pm0.02$  &  $1.02\pm0.01$  &  $0.13\pm0.02$  \\
\hline  
\end{tabular}
\end{table}

%
The sophisticated dust/star galaxy model of Popescu et al.\ (2000) and Tuffs
et al.\ (2004) incorporates clumpiness and an explicit treatment of various
grain compositions and sizes within a 3D distribution.  Solving the radiative
transfer equations they are able to measure the radial dependence of
attenuation in various wavebands.  M\"ollenhoff et al.\ (2006) has used this
to provide figures and tables for the effects of varying opacity and 
inclination on the parameters $\mu_0$ and $h$.  This was done using a range of
galaxy models with different integer values for the central, face-on
$B$-band opacity $\tau_B^f$.  

From a study of 10,095 nearby ($0.013 < z < 0.18$) 
galaxies in the Millennium Galaxy Catalogue, Driver et
al.\ (2007) have recently constrained this mean\footnote{Possible variations of 
opacity with Hubble Type, and thus also luminosity, are not yet firmly 
established.  
While Driver et al.\ (2007) and Shao et al.\ (2007) did not find evidence 
for such a variation, this is still an open issue (e.g.\ Masters et al.\ 2003) 
and needs to be addressed when 
far-infrared and sub-mm data become available for galaxies selected from 
optical surveys.} 
opacity to be $\tau_B^f = 3.8 \pm 0.7$.  
We explicitly note that this is a statistical determination of the opacity
and there may well be galaxy-to-galaxy variations outside of these bounds. 
It should also be noted that this value does not make spiral galaxies
optically thick in their outer parts or interarm regions (Popescu \& Tuffs
2007). Such high central opacities are however required, or rather consistent
with, the observed amount of infrared and sub-millimetre flux coming from the
thermally heated dust in galaxies (Driver et al.\ 2008).  Using the
above value for the opacity, and a linear
extrapolation between the gridded data points in M\"ollenhoff et al.,
the points in Figure~\ref{fig-Dust} show the combined effect of dust and viewing angle
(i.e.\ inclination) on the disc scalelength and central surface
brightness\footnote{We have used columns 4 and 5 from the tables in
M\"ollenhoff et al.\ (2006).} in the $B, V, I, J$ and $K$ passbands.  
The shaded region shows the range in behaviour at fixed inclination as $\tau_B^f$
changes from 3.1 to 4.5.

Using the linear regression routine FITEXY from Press et al.\ 
(1992)\footnote{The uncertainty on the independent variable, namely the
  inclination, was set to zero.  The uncertainty on the dependent variable
  ($\mu_{\rm 0,obs}$ or $h{\rm obs}$) 
  arose from the uncertainty of $\pm0.7$ on the value $\tau_B^f = 3.8$ (Driver
  et al.\ 2007).} two empirical relations have been fit to the simulated data points in
Figure~\ref{fig-Dust}.  
They are such that 
\begin{equation}
\mu_{\rm 0,obs} - \mu_{\rm 0,intrin} = 
a_{\lambda} + b_{\lambda}\left[ 2.5\log(\cos\, i) \right], 
\label{Eq_mu}
\end{equation}
and 
\begin{equation}
h_{\rm obs}/h_{\rm intrin} = 
c_{\lambda} - d_{\lambda} \log(\cos\, i), 
\label{Eq_h}
\end{equation}
for some inclination $i$, with $i=0$ degrees describing a face-on orientation.  
To
assist with the understanding of Figure~\ref{fig-Dust}, we note that in the
absence of dust, the line-of-sight depth through a disc increases as the disc
is inclined, such that the pathlength changes by $(\cos i)^{-1}$ to give 
$\mu_{\rm 0,obs} = \mu_{\rm 0,intrin} + 2.5\log(\cos\, i)$.
The parameters 
$a_{\lambda}$, $b_{\lambda}$, $c_{\lambda}$ and $d_{\lambda}$ in
equations~\ref{Eq_mu} and \ref{Eq_h} 
are a function
of wavelength, and given in Table~\ref{Tab_dust}.  These equations can be used
to correct observed values to the intrinsic face-on (dust-free) values.  
However, we caution that because the data in M\"ollenhoff et al.\ only 
extends to inclinations 
where $1-\cos i < 0.7$, the curves shown in Figure~\ref{fig-Dust} at 
higher inclinations are an extrapolation of these empirical models and may
therefore not be reliable for systems more inclined than 73 degrees.


\subsection{Bulge and disc magnitudes}

In addition to the above expressions, 
we shall also use the equations in Driver et al.\ 
(2008)
to correct the observed disc and bulge magnitudes for the effects of inclination and dust. 
These are given by 
\begin{equation}
M_{\rm disc, obs} - M_{\rm disc, intrin}  =  
b_1 + b_2 \left[ 1-\cos(i)\right] ^{b_3}, 
\label{Eq_Md}
\end{equation}
and
\begin{equation}
M_{\rm bulge, obs} - M_{\rm bulge, intrin} = 
d_1 + d_2 \left[ 1-\cos(i)\right] ^{d_3}. 
\label{Eq_Mb}
\end{equation}
where the coefficients in the above equations depend on wavelength and are 
provided in Driver et al.\ (2008) for various passbands, including the 
$B, r^{\prime}, i^{\prime}$ and $K$ bands.  We have used their $r^{\prime}$ and
$i^{\prime}$ coefficients for our $R$ and $I$ band data.  This approximation 
will not introduce any noticeable bias, as inspection of Figure~2 from 
Driver et al.\ (2008) will reveal.  

The effect of dust on bulges is far more severe than generally realised.  Past
studies of the $B/T$ ratio, while usually correcting the observed disc flux
for inclination and dust (but typically overlooking any attenuation once
adjusted to a face-on orientation), normally assume that the bulge is not
affected by dust.  This is wrong.  Moreover, given that bulges and discs have
their own inclination-attenuation correction, which is not surprising, there
is no single {\it galaxy} inclination-attenuation correction.  That is, one
must perform a bulge/disc separation.  Unfortunately this may severely hamper
the ability of large-scale studies (which have fitted single S\'ersic models
to two-component disc galaxies) to recover intrinsic galaxy parameters.

In this study we use the above formula to provide dust-corrected bulge
magnitudes and dust-corrected bugle-to-disc flux ratios.  However the
influence of dust on the individual S\'ersic parameters of the bulge
($\mu_{\rm e}, R_{\rm e}$ and $n$, see equation~\ref{EqSB}) is not well known.
It will of course depend not only on the wavelength of one's observations but
also on the precise star/dust geometry,
with concentrated high-$n$ bulges likely experiencing a greater degree of
attenuation.  Moreover, the overwhelming majority of optical bulge
parameters reported in the literature are dependent, to some degree, on the
happenstance inclination of their galaxy's disc.  As yet unknown corrections
to a dust-free configuration are required if we are to have an accurate set of
physical parameters for the bulges.  Nonetheless, for now, we have 
tabulated $K$-band S\'ersic bulge parameters as these will be the least 
affected by dust.

\section{The Galaxy Data} \label{SecData}

In this study we primarily investigate the properties of `late-type galaxies', 
and more generically `disc galaxies'.  Today, the term `early-type galaxy' is often
used to refer to both elliptical galaxies and lenticular galaxies.  While the
average properties of elliptical galaxies are not considered to be a function
of inclination --- and so one should identify and exclude such systems from
current inclination-dependent analyses of the effect of dust --- this is not
true for systems with large scale discs.  We have therefore included
lenticular disc galaxies in our analysis, rather than grouping/excluding them
with the elliptical galaxies in an early-type galaxy bin.

However we note that elliptical galaxies are certainly not always devoid of dust 
(e.g. Ebneter \& Balick 1985; Sadler \& Gerhard 1985; Ebneter, Djorgovski \&
Davis 1988; Leeuw et al.\ 2008).  Indeed, they also frequently possess 
small nuclear dust lanes and dusty nuclear discs (e.g.\ Ferrari et al.\ 1999; Rest
et al.\ 2001) which probably originate from sources such as type Ia SNe and
stellar mass loss from, for example, the winds of red giant stars (Ciotti et
al.\ 1991; Calura et al.\ 2008).  On a galactic scale, however, the dust is
destroyed rather quickly by ion sputtering from the hot X-ray gas (e.g.\ Temi,
Brighenti \& Mathews 2007).  In contrast, lenticular galaxies, such as the
Sombrero galaxy, are commonly observed to possess considerable amounts of dust in
their discs.  In this regard lenticular galaxies are more like spiral
galaxies.  Interestingly, Bekki, Couch \& Yasuhiro (2002) and
Arag{\'o}n-Salamanca, Bedregal \& Merrifield (2006) have argued that
lenticular galaxies may be spiral galaxies in which the star formation has
been turned off and the spiral pattern dispersed.  If such a morphological
transformation occurs, one would simultaneously require the bulge-to-disc flux
ratio to increase.  This may proceed via the passive fading of the disc and,
as noted by Driver et al.\ (2008), upon removal of centrally located dust.

\subsection{Literature catalogues} \label{Sec_Cat}






There has been a number of papers which have provided structural parameters
for nearby disc galaxies.
At the end of 2005, 
when selecting which of these would be suitable for our 
compilation of bulge and disc parameters, 
we applied the following criteria: 
(i) a S\'ersic bulge model had been simultaneously fit with an
exponential disc model; 
(ii) roughly a couple of dozen or more galaxies had been modelled;
(iii) morphological (Hubble or T)\footnote{We used 
de Vaucouleurs' T-type classification from the RC3 catalogue 
(de Vaucouleurs et al.\ 1991) which is roughly such that: 
$-3\leq T\leq -1$ (S0), T=0 (S0/a), T=1 (Sa), T=2 (Sab) ...\ T=9 (Sm), T=10 (Irr).} 
types existed for the galaxies; 
(iv) the galaxies appeared large enough that their bulges were well-resolved (this
meant that we only used studies with galaxy redshifts typically 
less than 0.03--0.04);
(v) studies of early-type galaxies (i.e. E and S0) which had not checked if a
bulge-only fit was superior to a bulge plus disc fit were excluded as many of the
fitted discs (and hence the structural parameters) may be spurious; 
and (vi) studies in which the fitted disc frequently failed to coincide with
the outer light-profile were also excluded, which included studies which often had the
fitted bulge model contributing more flux than the disc model at the outer
light profile.  
This latter problem is illustrated in Graham (2001a, his figure~7) and results
in overly large S\'ersic indices, $B/T$ flux ratios and $R_{\rm e}/h$ size
ratios.  
This can happen when using a (signal-to-noise)-weighted bulge $+$ disc 
fitting routine on a galaxy which has additional nuclear components and, 
as noted by Laurikainen et al.\ (2007) and Weinzirl et al.\ (2008), when 
significant bars are present. 
Our final sample selection, while almost certainly not all inclusive
due to unintentionally over-looked references, is given in Table~\ref{TabLit}.
Unfortunately the optical data from Hernandez-Toledo, Zendejas-Dominguez \&
Avila-Reese (2007) and Reese et al.\ (2007) appeared too late for us to
include.

\begin{table}
\caption{Number of galaxies in each passband from the literature
data that we were able to use. 
{\bf G02}     =  Graham (2002), low surface brightness (LSB) galaxies only,
  supplemented with LSB galaxies from Graham \& de Blok (2001) for 
  which redshifts have since become available; 
{\bf M04}     =  M\"ollenhoff (2004); 
{\bf MCH03}   =  MacArthur, Courteau \& Holtzman (2003); 
{\bf G01}     =  Graham (2001a, 2003);
{\bf MH01}    =  M\"ollenhoff \& Heidt (2001); 
{\bf SJ98}    =  Seigar \& James (1998); 
{\bf BGP03}   =  Balcells, Graham \& Peletier (2003); 
{\bf KdJSB03} =  Knapen et al.\ (2003); 
{\bf GPP04}   =  Grosb{\o}l, Patsis \& Pompei (2004);
{\bf LSB05}   =  Laurikainen, Salo \& Buta (2005); 
{\bf DD06}    =  Dong \& De Robertis (2006); 
{\bf KdJP06}  =  Kassin, de Jong \& Pogge (2006). 
\label{TabLit} 
}
\begin{tabular}{lcccccl}
\hline
Study     &  $B$ &  $V$  &  $R$  &  $I$  &   $K$     \\  
\hline  
G02       &  31  &  20   &   37  &  21   &   ...     \\  
M04       &  25  &  25   &  24   &  25   &   ...     \\  
MCH03     &  56  &   38  &  60   &  ...  &  ...      \\  
G01       &  74  &  ...  &   72  &  68   &   74      \\  
MH01      &  ... &  ...  &  ...  &  ...  &   40      \\  
SJ98      &  ... &  ...  &  ...  &  ...  &   44      \\  
BGP03     &  ... &  ...  &  ...  &  ...  &   19      \\  
KdJSB03   &  ... &  ...  &  ...  &  ...  &   28$^a$  \\  
GPP04     & ...  &  ...  &  ...  &  ...  &   53      \\  
LSB05     &  ... &  ...  &  ...  &  ...  &   22$^a$  \\  
DD06      &  ... &  ...  &  ...  &  ...  &   113     \\  
KdJP06    &  ... &  ...  &  ...  &  ...  &   15$^a$  \\  
Total     & 186  &  83   &  193  &  114  &   408     \\  
\hline  
\end{tabular}

\noindent
$^a$ $K_s$ filter.
\end{table}

Individual studies meeting the above criteria contained on the order of $\sim$20 to
$\sim$100 galaxies.  Previously, when these galaxies had been binned into their
morphological type (e.g.\ Sab, Scd, etc.), this tended to result in less than
ideal numbers per bin in each of these individual studies.  As a result, past
answers to questions of a statistical nature, such as the typical value of,
and range in, some physical parameter for a given morphological type, were
poorly defined.  By combining here (see Section~\ref{SecHomo}) the data from
these studies, we hope to better answer such questions.  Importantly, we
also apply corrections to the
magnitudes of both the discs {\it and} the bulges (see
Section~\ref{Sec_dust}).

We stress that while the galaxies used here are typical representations of
galaxies along the Hubble sequence, we have applied no sample selection
criteria to provide a magnitude-limited sample.  Given the
inclination-dependent reduction to magnitudes due to dust, we note in passing
that such a task is not as simple as one may first think: after correcting for
dust, one's original magnitude-limited sample will subsequently have a rather
jagged boundary.  However the desire to include as many galaxies as possible,
rather than this issue, has been the driving ratioanle employed here.

\subsection{Homogenisation} \label{SecHomo}

We have started with the catalogues of S\'ersic-bulge and exponential-disc parameters
from the studies listed in Table~\ref{TabLit}. 
While the exponential model used to describe the radial stellar distribution 
in discs has been around for a long time (e.g.\ Patterson 1940; de
Vaucouleurs 1957; Freeman 1970), 
%
%
S\'ersic's (1963) 3-parameter model 
has only become fashionable over the last decade.  It can be written as 
\begin{equation}
\mu (R)=\mu_{\rm e} + 1.086b_n\left[\left(R/R_{\rm e}\right)^{1/n}-1\right], 
\label{EqSB}
\end{equation}
where $\mu_{\rm e} = -2.5\log I_{\rm e}$ is the surface brightness at the effective half-light radius
$R_{\rm e}$, and $n$ is the S\'ersic index quantifying the radial
concentration of the stellar distribution.  The quantity $b_n$ is not
a parameter but instead a function of $n$ such that $\Gamma(2n) =
2\gamma(2n,b_n)$, where $\Gamma$ and $\gamma$ are the complete and 
incomplete gamma functions as given in Graham \& Driver (2005). 
For $ 0.5 < n < 10, b_n \approx 1.9992n - 0.3271$ (Capaccioli 1989). 
The exponential model used to describe discs can be obtained by setting
$n=1$ in S\'ersic's model, however it is more commonly expressed as 
\begin{equation}
\mu(R) = \mu_0 + 1.086(R/h), 
\end{equation}
in which $\mu_0$ is the central ($R=0$) surface brightness and $h$ is the
exponential disc scalelength. 

The apparent magnitude of an exponential disc is such that 
\begin{equation}
m_{\rm disc}  = \mu_0 -2.5\log(2\pi h^2),
\label{Eq_disc}
\end{equation}
and that of a S\'ersic bulge is given by the expression 
\begin{equation}
m_{\rm bulge} = \mu_{\rm e} -2.5\log(2\pi R_{\rm e}^2)
-2.5\log\left[ \frac{n{\rm e}^b}{b^{2n}}\Gamma(2n) \right] ,
\label{Eq_bulge}
\end{equation}
with both $h$ and $R_{\rm e}$ in arcseconds, and $\mu_0$ and $\mu_{\rm
  e}$ in mag arcsec$^{-2}$.
The gamma function $\Gamma$ 
has the property $2n\Gamma(2n) = \Gamma(2n+1) = (2n)!$.
Conversion from apparent magnitudes $m$ to absoulte magnitudes $M$ proceeds via
the standard expression $m-M = 25 + 5\log({\rm Distance} [Mpc])$. 
Although not done here, applying one's preferred stellar mass-to-light ratio 
to the absolute luminosities yields the associated stellar masses. 

In practice, because we use the observed disc magnitudes, due to the apparent
ellipticity $\epsilon$ of these discs --- as they are seen in projection -- 
equation~\ref{Eq_disc} had the $h^2$ term replaced with $h^2(1-\epsilon)$.
The quantity $1-\epsilon$ is equal to the observed minor-to-major axis 
ratio of the disc. 
As we do not have information on the ellipticity/triaxiality of the bulges, 
this work has assumed they are spherical (but see M\'endez-Abreu et al.\ 2008). 
%
Unless specified in their respective papers, the inclination of each galaxy 
has been estimated using $\cos\, i = b/a$, where $b/a$ is the {\it observed}
outer axis ratio.  Discs of course have a finite thickness, and so the
applicability of this expression deteriorates when dealing with increasingly
edge-on galaxies.  In the past, discs have been modelled as transparent oblate
spheroids (e.g.\ Holmberg 1946; Haynes \& Giovanelli 1984; Guthrie 1992) 
giving rise to the relation $\cos^2 i = [(b/a)^2 - Q^2]/[1-Q^2]$, where $Q =
c/a$ is the {\it intrinsic} short-to-long axis ratio.  Transparent edge on
discs will have an ellipticity equal to Q rather than zero.
%
%
Not surprisingly, the above relation breaks down in the presence of 
obscuring dust (M\"ollenhoff et al.\ 2006. their section~4.3 and Fig.12 which 
plots $[b/a]_{\rm obs}/[\cos\, i]_{\rm true}$).
%

The recovery of disc inclinations is not a major problem here because the
majority of our galaxies have observed axis ratios $b/a > 0.34$ and therefore
inclinations less than $\sim$70 degrees, where the use of $\cos\, i = b/a$
remains a reasonable approximation.  Only one galaxy (UGC 728, $b/a=0.284$) in
our optical data has $b/a$ smaller than 0.34.  In our $K$-band sample of
over 400 galaxies, where the 
effect of dust on magnitudes and scalelengths are small, there are
three samples with (some) discs having $b/a < 0.34$.  These include (1) Dong \& De
Robertis (2006) where a mere six per cent of the galaxies have inclinations
greater than 75 degrees and (2) M\"ollenhoff \& Heidt (2001) from which only
six galaxies have $b/a < 0.34$.  As such, these few galaxies have no
statistically significant impact on the results.  However, for the 19
early-type disc galaxies from (3) Balcells et al.\ (2003), the observed
ellipticities reach as high as 0.83.  Unfortunately M\"ollenhoff et al.\ (2006) do not
provide corrections to $\cos\, i$ for inclinations greater than 73 degrees,
and so we have adopted the $K$-band value for $Q = c/a = 0.11$ which was used in the
dust model of Tuffs et al.\ (2004).\footnote{The average $B$-band value of $Q$
  from Xilouris et al.\ (1999), which was used by Tuffs et al.\ (2004), is
  0.074.  The intrinsic stellar scaleheight reported by Xilouris et al.\
  (1999) is largely independent of wavelength, while the intrinsic stellar
  scalelength decreases with wavelength.  The dust model of Tuffs et al.\
  (2004) used a $K$-band intrinsic stellar scalelength that was 0.683 times
  the $B$-band value, giving rise to the value 0.074/0.683 $\approx$ 0.11.}
While this is not ideal, although dust is less of an issue in the $K$-band,
we note that for the most edge-on disc galaxy, 
with an observed ellipticity equal to 0.83, 
the derived inclination changes from $\sim$80 degrees to 82.5 degrees, i.e.\ 
$\cos\, i$ only changes from 0.17 to 0.13.

From the papers listed in Table~\ref{TabLit} we took, when available or
derivable, the
five main structural parameters ($\mu_0, h, \mu_{\rm e}, R_{\rm e}, n$).  We
also took the inclination $i$ of the disc, which, as noted above, usually came
from the ellipticity $\epsilon$ of the outer isophotes such that $\epsilon = 1
- b/a = 1 - \cos i$).
Potential intrinsic disc ellipticities, i.e.\ non-circular shapes of face-on
discs (Rix \& Zaritsky 1995; Andersen et al.\ 2001; Barnes \& Sellwood 2003;
Padilla \& Strauss 2008), 
and lopsidedness (e.g.\ Kornreich et al.\ 1998; Bournaud et al.\ 2006;
Reichard et al.\ 2008), have not been measured for our galaxy sample and are
consequently ignored.  However given the intrinsic disc ellipticity has been
estimated to have a mean value of only $\sim$0.05, the uncertainty this
introduces to the corrected scalelengths via equation~\ref{Eq_h} is less than
1 percent, and the uncertainty on the central surface brightnesses obtained
via equation~\ref{Eq_mu} is less than 0.04 mag arcsec$^{-2}$.  Except in the 
$K$-band, one can see
from Figure~\ref{fig-Dust} that the uncertainty in the face-on opacity
will introduces a greater source of scatter. 

For the distances to the galaxies we consulted Tonry et al.\ (2001) in the
case of the lenticular galaxies, and used the (Virgo $+$ GA $+$ Shapley)
distances given in NED\footnote{NASA Extragalactic Database (NED,
  http://nedwww.ipac.caltech.edu).}, for the remainders.  As a consequence, a
Hubble constant of 73 km s$^{-1}$ Mpc$^{-1}$ has effectively been assumed.
This Hubble constant is also the value reported in (Blakeslee et al.\ 2002)
and is the halfway point between the two values reported in van Leeuwen et
al.\ (2007).

Inclination-corrections to the disc surface brightnesses which had been
applied in some papers were undone according the prescription in each paper.
The resultant `observed' disc surface brightnesses were then corrected to a
face-on, dust-free value using equation~\ref{Eq_mu}.  Scalelengths --- which
had not been adjusted for inclination --- were corrected using
equation~\ref{Eq_h}, and disc magnitudes were subsequently computed using
equation~\ref{Eq_disc}.

Disc magnitudes were also derived using a second approach, in which the
observed (uncorrected) magnitude was corrected using equation~\ref{Eq_Md}.
Similarly, the observed bulge magnitudes have been corrected here using
equation~\ref{Eq_Mb}.
Published S\'ersic bulge parameters are not corrected for dust; expressions to
do so do not exist.  We therefore report on only the $K$-band S\'ersic bulge
parameters\footnote{Differences between the $K_s$ filter and the $K$-band have
been ignored.} which are summarised in Table~\ref{Tab_K}.

Galactic extinction corrections from Schlegel et al.\ (1998), as listed in
NED, have been applied (if not already done so in the original paper).
Cosmological redshift dimming also reduces the galaxy flux slightly, so we
have applied an adjustment of $2.5\log(1+z)^4$ to the magnitudes and surface
brightnesses when not already done so.  
This is not a huge adjustment, for example, it amounts to 0.1
mag at a redshift corresponding to 7,000 km s$^{-1}$.
Given the proximity of the galaxy samples, we have not applied 
evolutionary nor $K$-corrections.

In what follows we have grouped all S0 galaxies ($-3\leq T\leq -1$) 
into a single bin, denoted by $T=-1$ in our Figures and Tables.

\section{Structural parameters and their ratios}  \label{Sec_Single}

\subsection{The near-infrared}

\begin{figure}
\includegraphics[angle=270,scale=1.1]{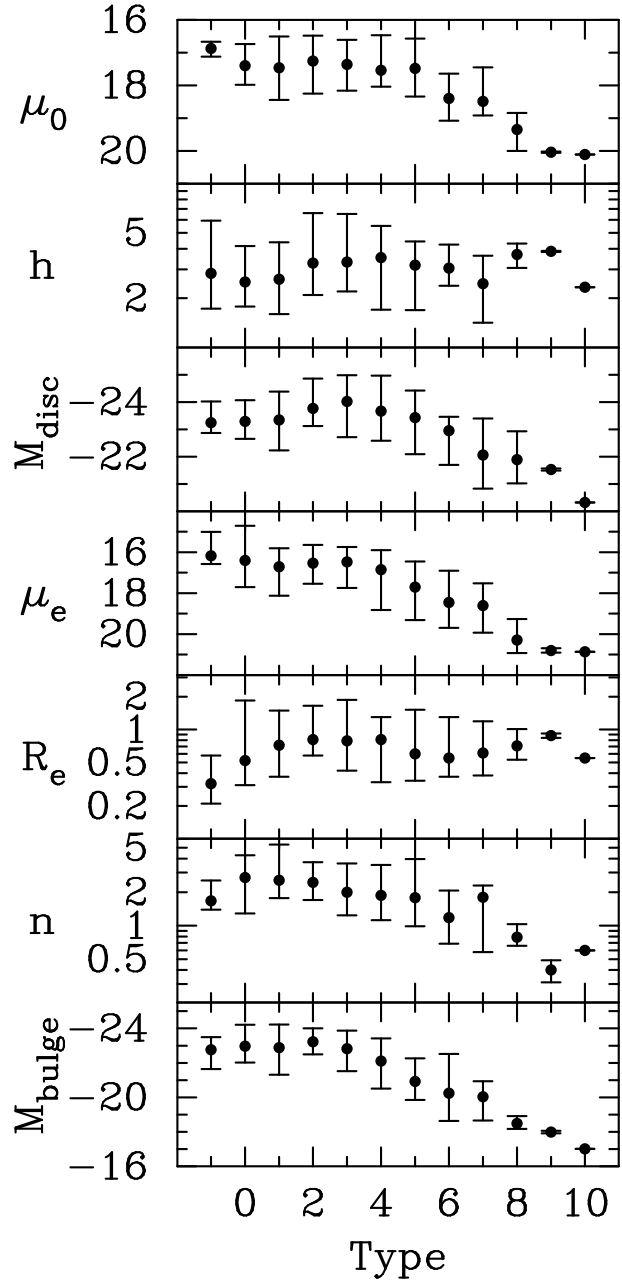}
\caption{
Standard ($K$-band) disc and bulge structural properties as a function of galaxy Type. 
For the disc, $\mu_0, h$ and $M_{\rm disc}$ are the central surface
brightness, scale-length and absolute magnitude. 
For the bulge, $\mu_{\rm e}, R_{\rm e}, n$ and $M_{\rm bulge}$ are the
effective surface brightness, effective half-light radius, S\'ersic index and 
absolute magnitude.
The data have been taken from Table~\ref{Tab_K}, except for the bulge
magnitudes which have been taken from Table~\ref{Tab_Ratio}. 
For each T-Type, the median is marked with a circle and the
`error bars' denote the 16 and 84 per cent quartiles of the full 
distribution, rather than uncertainties on the median value.
}
\label{figker}
\end{figure}

We have computed the median (and width of the central 68 per cent) from the
distributions of various structural parameters as a function of galaxy type.
In Table~\ref{Tab_K} we report the face-on, dust-free, $K$-band disc central
surface brightness $\mu_0$ and scalelength $h$, obtained using
equations~\ref{Eq_mu} and \ref{Eq_h}.  These values of $\mu_0$ and $h$ have
been used to derive, via equation~\ref{Eq_disc}, the disc magnitudes given in
Table~\ref{Tab_K}; they are also shown in Figure~\ref{figker}.
%
%
Spiral galaxies are commonly referred to as `early' type or `late' type 
(Hubble 1926)\footnote{Hubble (1926) referred to Sa galaxies as
  early-type, Sb as intermediate, and Sc as late-type.}.  
Given that morphological types are not always uniquely
assigned (Lahav et al.\ 1995), and often one only knows roughly what the
actual type is, we therefore report the above parameters and ratios for the following
three disc galaxy classes: 
lenticular galaxies (S0, S0/a); 
early-type spiral galaxies (Sa, Sab, Sb); and 
late-type spiral galaxies (Scd, Sd, Sm).
To help reduce cross contamination, the Sbc and Sc types are not used in our
galaxy class classification.

\begin{table*}
\caption{
$K$-band structural parameters of disc galaxies.
The median, $+/-$34 per cent of the distribution about the median, is
shown as a function of galaxy type (Column 1).
Columns 2 and 3: Disc central surface brightness and scalelength 
corrected using equations~\ref{Eq_mu} and \ref{Eq_h}, respectively.
Column 4: Disc magnitude computed using equation~\ref{Eq_disc} and the entries in column 2 and 3.
Columns 5, 6 and 7: Observed bulge effective surface brightness, radius, and S\'ersic index.
Column 8: Bulge magnitude computed using equation~\ref{Eq_bulge} and 
the entries in column 5--7 (no dust correction has been applied here). 
Column 9: Logarithm of the bulge luminosity density $\rho_{\rm e} \equiv \rho(r=R_{\rm e}) 
L_{\sun ,K}$ pc$^{-3}$.
Multiplying $\rho_{\rm e}$ by the appropriate $K$-band stellar mass-to-light ratio
will give the stellar mass density at $r=R_{\rm e}$ (see equation~\ref{r2_app}). 
The final three rows are such that $T\leq 0$ are the lenticular galaxies,
$T=1$-3 are the early-type spiral galaxies, and  $T=6$-9 are the late-type
spiral galaxies. 
\label{Tab_K}
}
\begin{tabular}{lcccccccc}
\hline
 & & & & & & & & \\
\hline
Type & $\mu_0$ & $h$ & $M_{\rm disc}$  &  $\mu_{\rm e}$ & $R_{\rm e}$ & $n$ & $M_{\rm bulge}$ & $\log \rho_{\rm e}$   \\
  & mag arcsec$^{-2}$ & kpc &    mag   & mag arcsec$^{-2}$ &  kpc     &     &   mag         &                 \\
1 &      2            & 3   &     4    &      5            &   6      &  7  &    8          &   9             \\
\hline
\multicolumn{9}{c}{Morphological Type} \\
$-3,-2,-1$, S0 & 
           $16.87^{+0.25}_{-0.20}$  &  $2.83^{+3.10}_{-1.10}$  &  $-23.24^{+0.37}_{-0.78}$  &  $16.18^{+0.40}_{-1.17}$  &  $0.32^{+0.26}_{-0.11}$  &  $1.67^{+0.87}_{-0.28}$  &  $-22.22^{+0.92}_{-0.90}$  & $0.78^{+0.27}_{-0.60}$ \\
 0, S0/a & $17.40^{+0.58}_{-0.66}$  &  $2.51^{+1.65}_{-0.73}$  &  $-23.29^{+0.64}_{-0.78}$  &  $16.40^{+1.31}_{-1.69}$  &  $0.52^{+1.32}_{-0.21}$  &  $2.71^{+1.58}_{-1.42}$  &  $-22.58^{+0.67}_{-1.51}$  & $0.13^{+1.06}_{-0.89}$ \\
 1, Sa   & $17.46^{+0.98}_{-0.95}$  &  $2.61^{+1.77}_{-1.01}$  &  $-23.34^{+1.11}_{-1.04}$  &  $16.71^{+1.42}_{-0.90}$  &  $0.72^{+0.77}_{-0.35}$  &  $2.56^{+2.79}_{-0.79}$  &  $-22.67^{+1.52}_{-1.39}$  & $0.11^{+0.46}_{-0.85}$ \\
 2, Sab  & $17.26^{+0.99}_{-0.78}$  &  $3.27^{+3.34}_{-1.18}$  &  $-23.77^{+0.65}_{-1.09}$  &  $16.54^{+1.00}_{-0.89}$  &  $0.81^{+0.84}_{-0.23}$  &  $2.45^{+1.27}_{-0.75}$  &  $-23.11^{+0.95}_{-0.78}$  & $0.20^{+0.44}_{-0.77}$ \\
 3, Sb   & $17.36^{+0.80}_{-0.75}$  &  $3.32^{+3.20}_{-1.12}$  &  $-24.02^{+1.31}_{-0.96}$  &  $16.48^{+1.27}_{-0.73}$  &  $0.79^{+1.08}_{-0.37}$  &  $2.00^{+1.62}_{-0.76}$  &  $-22.59^{+1.29}_{-1.09}$  & $0.17^{+0.51}_{-0.75}$ \\
 4, Sbc  & $17.54^{+0.50}_{-1.07}$  &  $3.54^{+1.98}_{-1.84}$  &  $-23.67^{+1.09}_{-1.30}$  &  $16.86^{+1.96}_{-0.96}$  &  $0.81^{+0.49}_{-0.48}$  &  $1.87^{+1.64}_{-0.75}$  &  $-21.88^{+1.52}_{-1.34}$  & $0.04^{+0.62}_{-0.80}$ \\
 5, Sc   & $17.48^{+0.86}_{-0.91}$  &  $3.18^{+1.26}_{-1.49}$  &  $-23.43^{+1.34}_{-0.99}$  &  $17.71^{+1.61}_{-1.25}$  &  $0.60^{+0.92}_{-0.26}$  &  $1.78^{+2.18}_{-0.79}$  &  $-20.72^{+1.07}_{-1.41}$  & $-0.27^{+0.73}_{-0.78}$ \\
 6, Scd  & $18.40^{+0.68}_{-0.76}$  &  $3.05^{+1.19}_{-0.67}$  &  $-22.95^{+1.25}_{-0.51}$  &  $18.46^{+1.24}_{-1.55}$  &  $0.55^{+0.75}_{-0.18}$  &  $1.18^{+0.89}_{-0.49}$  &  $-20.13^{+1.61}_{-2.28}$  & $-0.43^{+0.79}_{-0.47}$ \\
 7, Sd   & $18.49^{+0.43}_{-1.04}$  &  $2.45^{+1.18}_{-1.03}$  &  $-22.06^{+1.23}_{-1.34}$  &  $18.61^{+1.32}_{-1.09}$  &  $0.61^{+0.58}_{-0.23}$  &  $1.80^{+0.49}_{-1.22}$  &  $-19.93^{+1.39}_{-0.88}$  & $-0.33^{+0.30}_{-1.06}$ \\
 8, Sdm  & $19.35^{+0.65}_{-0.51}$  &  $3.69^{+0.61}_{-0.63}$  &  $-21.89^{+0.87}_{-1.04}$  &  $20.29^{+0.64}_{-1.02}$  &  $0.71^{+0.30}_{-0.18}$  &  $0.79^{+0.24}_{-0.13}$  &  $-18.35^{+0.32}_{-0.43}$  & $-1.28^{+0.50}_{-0.44}$ \\
 9, Sm   & $20.04^{+0.02}_{-0.02}$  &  $3.86^{+0.03}_{-0.03}$  &  $-21.53^{+0.04}_{-0.04}$  &  $20.80^{+0.11}_{-0.11}$  &  $0.88^{+0.04}_{-0.04}$  &  $0.40^{+0.09}_{-0.09}$  &  $-17.85^{+0.06}_{-0.06}$  & $-1.61^{+0.06}_{-0.06}$ \\
10, Irr  & $20.11^{+0.00}_{-0.00}$  &  $2.33^{+0.00}_{-0.00}$  &  $-20.32^{+0.00}_{-0.00}$  &  $20.87^{+0.00}_{-0.00}$  &  $0.55^{+0.00}_{-0.00}$  &  $0.60^{+0.00}_{-0.00}$  &  $-16.90^{+0.00}_{-0.00}$  & $-1.44^{+0.00}_{-0.00}$ \\
\multicolumn{9}{c}{Morphological Class} \\                                                                                                                                                                                         
$-3\leq T\leq 0$ &                                                                                                                                                                                                                  
           $17.21^{+0.74}_{-0.49}$  &  $2.60^{+1.63}_{-0.83}$  &  $-23.27^{+0.60}_{-0.81}$  &  $16.32^{+1.10}_{-1.51}$  &  $0.51^{+1.02}_{-0.26}$  &  $2.08^{+1.76}_{-0.74}$  &  $-22.54^{+0.78}_{-1.47}$  & $0.36^{+0.76}_{-0.99}$ \\
$T=1$-3  & $17.33^{+0.90}_{-0.84}$  &  $3.05^{+2.96}_{-1.14}$  &  $-23.73^{+1.21}_{-1.11}$  &  $16.55^{+1.20}_{-0.87}$  &  $0.78^{+0.81}_{-0.34}$  &  $2.18^{+1.82}_{-0.77}$  &  $-22.73^{+1.43}_{-1.18}$  & $0.15^{+0.50}_{-0.78}$ \\
$T=6$-9  & $18.49^{+1.02}_{-0.87}$  &  $3.05^{+1.03}_{-0.94}$  &  $-22.58^{+1.59}_{-0.72}$  &  $18.62^{+1.42}_{-1.73}$  &  $0.60^{+0.62}_{-0.36}$  &  $1.15^{+0.95}_{-0.55}$  &  $-19.68^{+1.76}_{-1.43}$  & $-0.61^{+0.56}_{-0.80}$ \\
\hline
\end{tabular}
\end{table*}

\begin{figure}
\includegraphics[angle=270,scale=0.45]{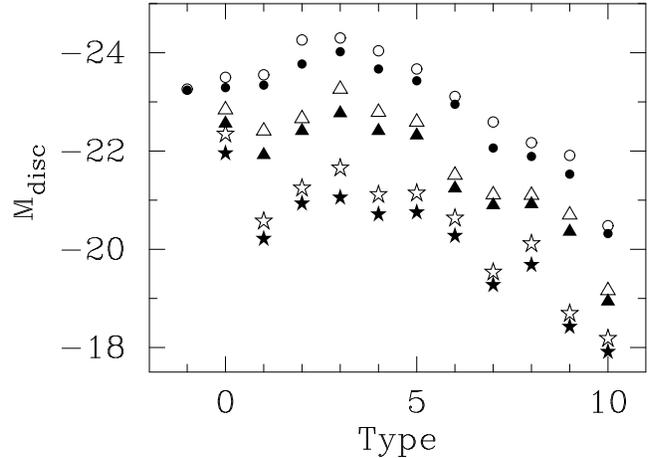}
\caption{
Comparison of our two estimates of the (corrected) disc magnitudes.
The filled symbols show the median values obtained using the corrections in 
equations~\ref{Eq_mu} and \ref{Eq_h}, while the open symbols show the
median values obtained using equation~\ref{Eq_Md}. 
$B$-band: stars. 
$I$-band: triangles.
$K$-band: circles. 
}
\label{fig-Mag-disc}
\end{figure}

In Table~\ref{Tab_Ratio} one can find the dust-corrected disc and bulge
magnitudes obtained using equations~\ref{Eq_Md} and \ref{Eq_Mb}, respectively.
While equation~\ref{Eq_Mb} was applied to the bulge magnitudes in
Table~\ref{Tab_K}, equation~\ref{Eq_Md} was not applied to the disc magnitudes
from this table but the disc magnitudes obtained using the observed disc
scalelengths and central surface brightnesses.
%
%
Given that we now have two estimates of the corrected disc magnitude, one in
Table~\ref{Tab_K} and the other in Table~\ref{Tab_Ratio}, these are compared
in Figure~\ref{fig-Mag-disc} and shown to agree within a couple of tenths of a
magnitude or better.  A change of 0.2 mag for the disc magnitude corresponds
to a change in $\log (B/D)$ of 0.08.  We are also able to show such a
comparison for the $B$- and $I$-band, as both sets of dust corrections are
available in these bands.
The slight difference in Figure~\ref{fig-Mag-disc} is thought to arise from
the exponential function not providing a perfect description of dusty discs
(see M\"ollenhoff et al.\ (2006, their Figure~2). Indeed, due to the prevalence
of centrally located dust, the best-fitting exponential model can be seen in
M\"ollenhoff et al.\ to overestimate the observed flux at small radii.
Therefore, while the corrections between the measured and intrinsic central
surface brightness and disc scalelength given in equations~\ref{Eq_mu} and
\ref{Eq_h} are appropriate, application of equation~\ref{Eq_disc}, using
$\mu_{\rm 0,intrin}$ from equation~\ref{Eq_mu}, and $h_{\rm intrin}$ from
equation~\ref{Eq_h}, can lead to a slight over-estimate of the actual observed
disc magnitude.  As a result, the corrections to the observed disc magnitude
using this approach are not quite as large as they should be (C.Popescu \& R.Tuffs
2008, priv.\ comm.).

\subsubsection{The bulge-to-disc size ratio $R_{\rm e}/h$}

\begin{table*}
\caption{
$K$ band parameters of disc galaxies. 
The median, $\pm$68/2 per cent of the distribution on either side of the median, is
shown as a function of galaxy type (Column 1).
Column 2: Bulge-to-disc size ratio using columns 6 and 3 from Table~\ref{Tab_K}
Column 3: Bulge magnitude, obtained from the observed flux, corrected using equation~\ref{Eq_Mb}. 
Column 4: Disc magnitude, obtained from the observed flux, corrected using equation~\ref{Eq_Md}. 
Column 5: Bulge-to-disc flux ratio using columns 2 and 3.
Column 6: Number of data points. 
The bulge-to-total flux ratio can be obtained from 
the expression $B/T = [1+(D/B)]^{-1}$, where $B/D$ is the bulge-to-disc flux ratio.
\label{Tab_Ratio} 
}
\begin{tabular}{lccccc}
\hline
Type  &  $R_{\rm e}/h$  &  $M_{\rm bulge}$  &  $M_{\rm disc}$  &  $\log(B/D)$  & \#  \\
\hline
\multicolumn{6}{c}{Morphological Type} \\
$-3,-2,-1 $, S0                            
          & $0.12^{+0.07}_{-0.04}$ & $-22.76^{+1.12}_{-0.74}$  &  $-23.26^{+0.40}_{-1.28}$  &  $-0.54^{+0.28}_{-0.27}$  &  16  \\ 
 0, S0/a  & $0.22^{+0.37}_{-0.09}$ & $-22.97^{+0.95}_{-1.24}$  &  $-23.50^{+0.48}_{-0.59}$  &  $-0.31^{+0.46}_{-0.24}$  &  30  \\ 
 1, Sa    & $0.31^{+0.20}_{-0.17}$ & $-22.89^{+1.58}_{-1.34}$  &  $-23.55^{+1.04}_{-1.05}$  &  $-0.34^{+0.40}_{-0.32}$  &  45  \\ 
 2, Sab   & $0.24^{+0.22}_{-0.10}$ & $-23.23^{+0.73}_{-0.78}$  &  $-24.26^{+0.92}_{-0.92}$  &  $-0.54^{+0.53}_{-0.41}$  &  38  \\ 
 3, Sb    & $0.21^{+0.15}_{-0.07}$ & $-22.83^{+1.31}_{-1.04}$  &  $-24.30^{+1.15}_{-0.87}$  &  $-0.60^{+0.28}_{-0.49}$  &  60  \\ 
 4, Sbc   & $0.21^{+0.11}_{-0.09}$ & $-22.10^{+1.59}_{-1.32}$  &  $-24.04^{+1.14}_{-1.29}$  &  $-0.82^{+0.28}_{-0.42}$  &  79  \\ 
 5, Sc    & $0.22^{+0.27}_{-0.09}$ & $-20.92^{+1.07}_{-1.35}$  &  $-23.67^{+1.19}_{-1.07}$  &  $-1.06^{+0.43}_{-0.34}$  &  94  \\ 
 6, Scd   & $0.19^{+0.10}_{-0.06}$ & $-20.24^{+1.60}_{-2.28}$  &  $-23.11^{+1.30}_{-1.03}$  &  $-1.23^{+0.75}_{-0.28}$  &  28  \\ 
 7, Sd    & $0.24^{+0.07}_{-0.10}$ & $-20.04^{+1.38}_{-0.90}$  &  $-22.59^{+1.39}_{-1.04}$  &  $-1.06^{+0.16}_{-0.50}$  &  11  \\ 
 8, Sdm   & $0.19^{+0.14}_{-0.07}$ & $-18.49^{+0.32}_{-0.43}$  &  $-22.17^{+0.85}_{-1.07}$  &  $-1.49^{+0.36}_{-0.36}$  &   4  \\ 
 9, Sm    & $0.23^{+0.01}_{-0.01}$ & $-17.99^{+0.07}_{-0.07}$  &  $-21.91^{+0.10}_{-0.10}$  &  $-1.57^{+0.01}_{-0.01}$  &   2  \\ 
10, Irr   & $0.24^{+0.00}_{-0.00}$ & $-17.01^{+0.00}_{-0.00}$  &  $-20.48^{+0.00}_{-0.00}$  &  $-1.39^{+0.00}_{-0.00}$  &   1  \\ 
\multicolumn{6}{c}{Morphological Class} \\                                                            
$-3\leq T \leq 0$                                                                                     
          & $0.17^{+0.24}_{-0.08}$ & $-22.96^{+1.00}_{-1.18}$  &  $-23.49^{+0.51}_{-0.63}$  &  $-0.37^{0.40}_{-0.35}$  &  46 \\ 
$T=1$-3   & $0.24^{+0.22}_{-0.10}$ & $-22.95^{+1.38}_{-1.16}$  &  $-24.15^{+1.15}_{-0.95}$  &  $-0.51^{0.45}_{-0.37}$  & 143 \\ 
$T=6$-9   & $0.22^{+0.08}_{-0.09}$ & $-19.80^{+1.72}_{-1.45}$  &  $-22.87^{+1.49}_{-1.14}$  &  $-1.19^{0.43}_{-0.40}$  &  45 \\ 
\hline
\end{tabular}
\end{table*}

Using exponential bulge models, de Jong (1996b) suggested that the bulge-to-disc
size ratio was independent of Hubble type.  This unexpected result was
reiterated by 
Courteau, de Jong \& Broeils (1996) using 
an additional $\sim$250 Sb/Sc galaxies imaged in the $r$-band. 
%
%
Adding to the mystery, Graham \& Prieto (1999) pointed 
out that de Jong's early-type spiral galaxies actually had 
an average $R_{\rm e}/h$ ratio --- derived using his exponential bulge
models --- 
that was {\it smaller} (at the $3\sigma$ level in the $K$-band) than
the average $R_{\rm e}/h$ ratio from his late-type spiral galaxies.

%
Rather than only using the effective radii from de Jong's exponential bulge
models, Graham \& Prieto explored use of the $R_{\rm e}$ values from de Jong's
``best-fitting'' bulge model; which was either an $R^{1/4}$ model, an
$R^{1/2}$ model or an exponential $R^{1/1}$ model.  Doing so, the mean $R_{\rm
e}/h$ ratio was shown to be slightly larger for the early-type disc galaxies
than the late-type disc galaxies.  Following up on this, Graham (2001a) fitted
S\'ersic bulge (plus exponential disc) models to all of the light profiles
used by de Jong.  Graham revealed little difference between the early- and
late-type disc galaxy bulge-to-disc size ratios, and therefore ultimately
reached the same conclusion as Courteau et al.\ (1996) but on somewhat
different grounds.  To explain this apparent discrepancy with what one sees
when looking at spiral galaxies of different type, i.e.\ that early-type
spiral galaxies {\it appear} to have larger bulge-to-disc size ratios than
late-type spiral galaxies, Graham (2001a, his figure~21) subsequently
presented the ice-berg' model in which the intensity of the bulge varies
(is effectively raised or lowered) relative to the intensity of the disc. 

\begin{figure}
\includegraphics[angle=0,scale=0.44]{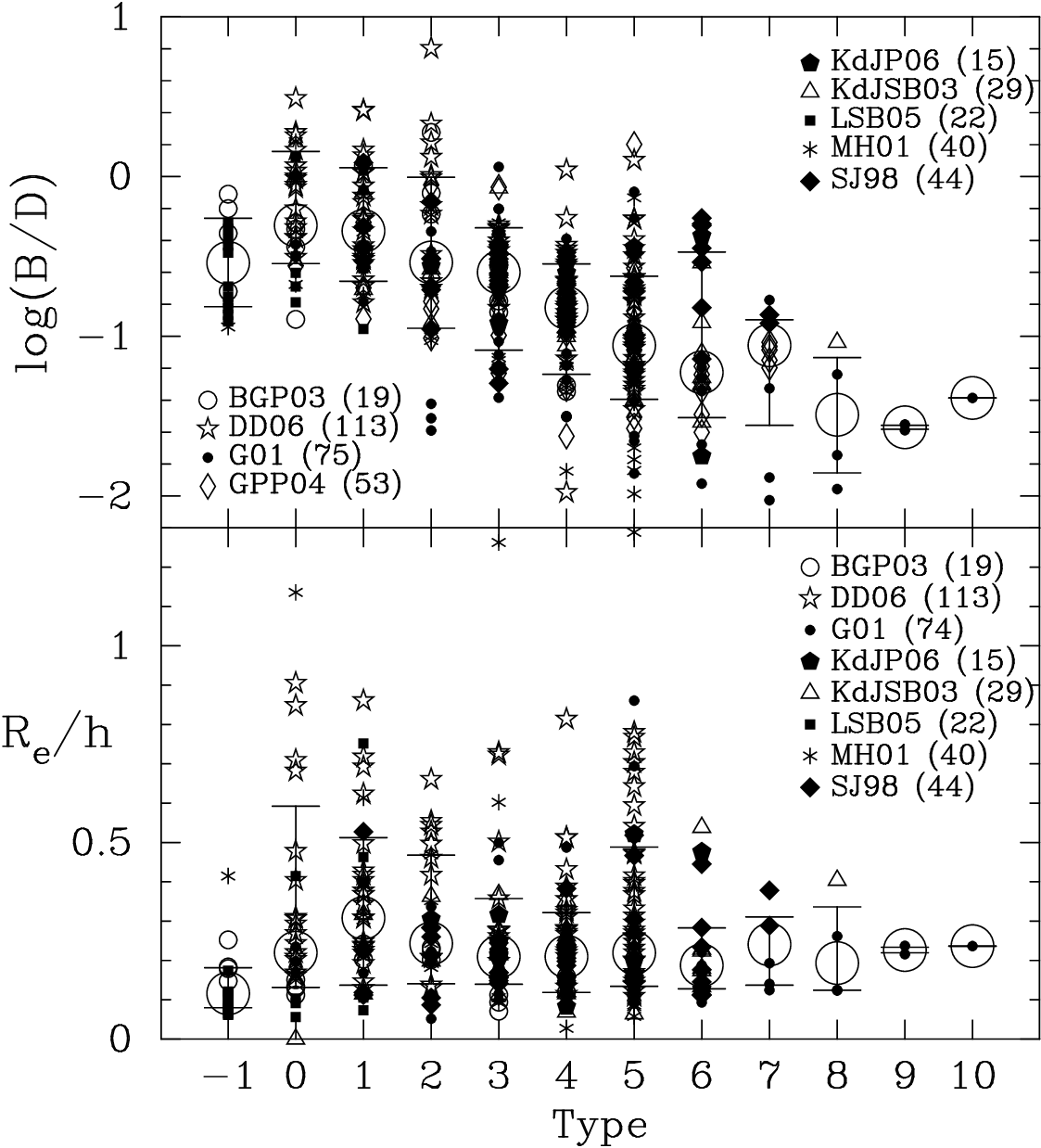}
\caption{
Top panel: $K$-band, bulge-to-disc size ratio as a function of galaxy Type.
Bottom panel: Logarithm of the $K$-band, bulge-to-disc flux ratio as a 
function of galaxy Type.  Values have been taken from Tables~\ref{Tab_Ratio}. 
For each Type, the median value is marked with a large circle and the
`error bars' denote the 16 and 84 per cent quartiles of the distribution; 
they thus enclose the central 68 per cent of each distribution. 
}
\label{fig-K-Ratio}
\end{figure}

Using $\sim$400 disc galaxies, 
Table~\ref{Tab_Ratio} and Figure~\ref{fig-K-Ratio}, present the $K$-band $R_{\rm
  e}/h$ size ratio as a function of disc galaxy type.  One of the nice
features of this data set is in fact its heterogeneous nature.  As a result,
possible biases in any one paper's modelling are minimised and/or effectively
cancelled by those from the other papers.  The results shown here are
therefore something of a consensus from many papers.  One can see that the
median value is rather stable at around $0.22^{+0.02}_{-0.03}$, with the only
departure from this being (i) an increase to 0.31 for the Sa galaxies and (ii)
a value of 0.12 for T-types less than zero.  
The first result arises from some of the shallow 2MASS data analysed by Dong
\& De Robertis (2006), which is responsible for the scatter to higher $R_{\rm
  e}/h$, and also higher $\log B/D$, ratios than displayed by the other data.
Excluding this data set, the remaining Sa galaxies have $R_{\rm e}/h = 0.25$
with a variance in this distribution of 0.04.  The second result above is
unusual: the size ratio is equal to or less than the lower 1$\sigma$ limit
from every other galaxy type (not to mention that lenticular galaxies are
commonly thought to have the largest bulge-to-disc ratios).  This result is
however again solely due to the data from one study.  Modelling the bulge and
bare (and disc) as separate components, Laurikainen et al.\ (2005) obtained
these smaller ratios.  In contrast, the barless galaxies from Balcells et al.\ (2003) with
T-type less than 0 have $\langle R_{\rm e}/h \rangle \sim 0.2$.  From the
$K$-band panel in Figure~\ref{fig-Dust}, one can see that uncertainties in the
disc inclination for the Balcells et al.\ (2003) sample will not change this
result by much.
%

We are unable to include the lenticular galaxies from Barway et al.\ (2007) as
neither their fits nor bulge/disc parameters have yet been released.  These
authors claim that the $R_{\rm e}/h$ ratio is not (roughly) constant for all
disc galaxies and that luminous lenticular galaxies can have values as high
as 5 to 10, implying the presence of small embedded discs whose light peters
out well before the half-light radius of the bulge component.
Such potentially new objects with bulge-to-disc size ratios $\sim$50 times
greater than the median value of regular lenticular galaxies studied by others
would surely be an important clue to the transition between disc and
elliptical galaxies.
If real, these discs might be more akin to the nuclear discs seen in many
elliptical galaxies (e.g.\ Rest et al.\ 2001).  However, 
%
%
we note that the sky-background should always
be derived independently of one's fitted models.  Barway et al.'s
(undesirable) treatment of the sky-background as a free parameter when fitting
the bulge and disc models may be responsible for the extreme size ratios they
report.  Such an approach to modelling galaxies modifies the real surface
brightness profile, and may result in the occurrence of item {\it (vi)} in
Section~\ref{Sec_Cat}.

\subsubsection{The bulge-to-disc flux ratio}

The inclination- and dust-corrected $K$-band bulge-to-disc flux ratios, $B/D$, 
are provided 
in Table~\ref{Tab_Ratio} and shown in Figure~\ref{fig-K-Ratio} as a function
of galaxy type.  Unlike the size ratio, the flux ratio does indeed decrease
with increasing galaxy type.  The early-type spiral galaxies (Sa-Sb) have a
median value of $\log(B/D) = -0.49$ (i.e.\ $B/D=0.32$, or equivalently a
bulge-to-total flux ratio $B/T=0.24$), where $B$ and $D$
represent the luminosity (not magnitude) of the bulge and disc respectively.
The late-type spiral galaxies (Scd-Sm) have a median value of $\log(B/D) =
-1.40$ ($B/D=0.04, B/T=0.04$).

The tabulated $K$-band $B/D$ flux ratios for the 12 lenticular galaxies in
Andredakis et al.\ (1995. their table~4) give a mean $B/T$ ratio of 0.28.
Similarly, as reported in Balcells et al.\ (2007), 
the mean (plus or minus the standard deviation) of the $B/T$ flux
ratio for the lenticular galaxies analysed in Balcells et al.\ (2003) is 0.25
($\pm0.09$).
Laurikainen et al.\
(2005) also obtained a similar result from their $K$-band data, reporting a
mean value of 0.24 ($\pm0.11$) from a sample of 14 S0 galaxies
while the data in Gadotti (2008) yields an $R$-band value of 0.28 from 
their 7 lenticular galaxies.  Our result implies 
that (50+68/2=) 84 percent of lenticular galaxies have $B/T$ ratios smaller
than $\sim$1/3.
The remark in Kormendy (2008) that ``almost no pseudobulges have $B/T >
1/3$'' could thus be expanded to read ``most disc galaxies have $B/T < 1/3$''. 

The above ratios, obtained from the collective average of nine modern studies (see
Table~\ref{TabLit}), are smaller than those reported in the 1980s.  Indeed, many
late-type spiral galaxies have bulge fluxes which are less than 4 per cent of
their galaxy's total light.  
While Shields et al.\ (2008) claim that the T=6 (Scd) galaxy NGC~1042 may be
a bulgeless galaxy, its optical light profile in B\"oker et al.\ (2003, their
Figure~1)\footnote{The bump in B\"oker et al.'s (2003)  optical light profile
  for NGC~1042 at 
$4\pm2$ kpc is due to prominant inner spiral arms which cause a clear 
excess above the disc's underlying exponential light profile.} 
 together with its $K$-band light profile in Knapen et al.\ (2003), 
plus the diffuse glow of central light in the optical images,
reveals an excess of flux over the inner 2 kpc which is well above that
defined by the inward extrapolation of the outer exponential light 
profile.
Our corrected $K$-band bulge magnitude (and $B/T$ ratio) for MGC~1042 is 
$-19.13$ mag (0.054), implying a supermassive black hole mass of 
$3\times 10^6 M_{\sun}$ (Graham 2007), in agreement with the upper 
bound estimated by Shields et al.\ (2008). 
We therefore caution that it can be difficult when determining if a galaxy is 
actually bulgeless, which is a topic of particular interest at the low-mass
end 
of relations involving supermassive black hole masses (e.g.\ Satyapal et al.\
2007, 2008).  It is not yet determined if 
massive black can holes form (at any redshift) before or without a host bulge. 
Local galaxies may shed insight into this `chicken-egg' 
problem of whether the bulge or black hole formed first, if not in tandem.



Laurikainen et al.\ (2007, their figure~1) recently presented a set of $K$-
and $H$-band $B/T$ flux ratios versus galaxy type for a sample of lenticular
galaxies and the spiral galaxies from the Ohio State University Bright Galaxy
Survey (Eskridge et al.\ 2002).  Although that data has not been fully
corrected for dust attenuation --- which is not insignificant in the $H$-band
(e.g.\ Peletier \& Willner 1992; Driver et al.\ 2008) --- the trend which 
they show embodies the general behaviour observed here.
%
%
In a separate study, 
Dong \& De Robertis (2006) claim that their $K$-band flux ratio versus galaxy
type diagram (their fig.7) agrees with the original $B$-band results given by
Simien \& de Vaucouleurs (1986).  While roughly true, such an agreement should
not occur given the differing bulge and disc stellar populations (and hence
colours in the $B$- and $K$-band) for late-type disc galaxies.  Aside from
this problem, much of the agreement arises from the partial cancellation of
two significant errors: Simien \& de Vaucouleurs over-estimated their bulge
flux due to the use of $R^{1/4}$ models while at the same time they neglected
the dimming effects of dust.  Working with $R^{1/n}$-bulge models in the
$K$-band, Dong \& De Robertis (2006) overcame these issues.

Recent claims that the luminosity of the bulge, rather than the $B/D$ ratio,
is the driving force behind the trend seen in Figure~\ref{fig-K-Ratio} 
(e.g.\ Trujillo et al.\ 2002;
Scannapieco \& Tissera 2003; 
Balcells et al.\ 2007) reflect the results in Yoshizawa \& Wakamatsu (1975,
their figures~1 and 2) and echo the remarks in Ostriker (1977) and Meisels \&
Ostriker (1984) that the bulge luminosity, and ergo mass, may be a key
parameter which distinguishes galaxies.
In Figure~\ref{figker} we show, separately, the bulge and disc
magnitude as a function of galaxy T-type. 
One can see that the disc magnitude is roughly constant from T$=-1$ to T$=4$
and then it falls by a couple of magnitude upon reaching T$=9$ (Sdm galaxy). 
On the other hand, the bulge magnitude is roughly constant from T$=-1$ to
T$=3$, but then falls five magnitude by T$=9$.   We therefore confirm that
the bulge magnitude is indeed predominantly responsible for the trend between 
the bulge-to-disc luminosity and disc galaxy type. 
In passing we note that Hernandez \& Cervantes-Sodi (2006) have 
advocated that the spin parameter is the physical quantity which determines a
disc galaxy's Hubble Type (see also Foyle et al.\ 2008); implying that the
spin parameter must therefore be connected with the magnitude of the bulge.

Simien \& de Vaucouleurs (1986) presented a Figure showing the mean difference
between the bulge and galaxy magnitude as a function of morphological T-type.
Also shown in their Figure was the associated statistical uncertainty on their 
mean differences, which is of course much smaller than the actual
scatter about each mean.  They fitted a quadratic relation to this $\delta
M$-T data, which has been popular in semi-analytical studies in which
bulge-to-disc ratios are known but little
information about the spiral arms is available.  Although one could fit a
relation to the median $B/D$ ratios as a function of T-type in 
Figure~\ref{fig-K-Ratio}, we feel that this may be misleading.  For example,
if $\log (B/D) = -0.5$, the actual T-type could range from -1 (S0) to 6 (Scd),
while an equation would give only one value.  The range of values reminds us
that the bulge-to-disc flux ratio was never used as the primary criteria for
classifying disc galaxy morphology.
However, given the prominace of a bulge reflects the relative dominance of
certain construction processes, which may be different to those which generate
spiral patterns, then it may at times be desirable to employ such quantitative
measures without recourse to Hubble type.

Knowing the stellar mass-to-light ratio of the bulge and disc components
allows one to transform the bulge-to-total flux ratios, $B/T$, into stellar
mass ratios.  This aids comparison with cosmological simulations of galaxies. 
The stellar bulge-to-total mass ratio is given by $ \{ 1 +
  \chi [(B/T)^{-1}-1]\} ^{-1}$, where $\chi = (M/L)_{\rm disc}/(M/L)_{\rm bulge}$ is
the ratio of the disc and bulge stellar mass-to-light ratios, and 
$B/T$ is of course the bulge-to-total flux ratio.  If we use a
slightly extreme (low) value of $\chi = 1/2$ (see Bruzual \& Charlot 2003),
then for $B/T = 0.25$ (0.33) we have a bulge-to-total stellar mass ratio of
0.4 (0.5).

To date, most simulations have a tendency to produce bulges rather than
(pure) discs because of baryon angular momentum losses during merger
events (Navarro \& Benz 1991; Navarro \& White 1994; van den Bosch 2001; 
D'Onghia et al.\ 2006).  To avoid this issue, while still
maintaining the hierarchical merger trees for CDM halos, potentially
unrealistic amounts of feedback have been invoked in the past.  Early feedback could blow
much of the baryons out of high-$z$ galaxy discs for their late-time, and
necessarily post major-merger, re-accretion into some preferred plane and
thereby produce the disc-dominated galaxies observed in the local Universe
(see Abadi et al.\ 2003a).
From the $\sim$1000 spiral galaxies simulated by Croft et al.\ (2008), at $z=1$ 
only 3.5 per cent have a bulge-to-total stellar mass ratio smaller than 0.5.
Given that 68 per cent of real lenticular galaxies have $B/T$ smaller than
$\sim$1/3, these simulations are clearly inconsistent with what is observed in
the Universe.  The disc galaxy simulated by Abadi (2003b) has a
stellar bulge-to-total mass ratio of 0.71, and as such it also does not represent a
typical disc galaxy.  While the high mass and force resolution $\Lambda$CDM 
simulated spiral galaxy in Governato et al.\ (2004) has $B/T = 0.26$,
subsequent high-resolution simulations (e.g.\ D'Onghia et al.\ 2006) reveal
that resolution issues are not the key problem but rather some primary physics may
still be missing.  Current implementation of supernovae feedback, for example, 
has been shown to play a significant role in the evolution of simulated
galaxies (Okamoto et al.\ 2005; Governato et al.\ 2007; Scannapieco et al.\ 2008 and references therein). 
%


The current data may hopefully help determine how bulges should be assigned 
their mass in 
semi-analytical models.  Such studies build galaxies by a trial and error
process in which the model assumptions, parameters and processes are collectively 
tweaked until the simulations resemble real galaxies.  
For example, when a simulated disc becomes unstable
(Mo et al.\ 1998; Cole et al.\ 2000), Croton et al.\ (2006, their equation~22)
transfer enough disc mass to the bulge in order to maintain stability, while
Bower et al.\ (2006, their equation~1) transform the entire system into an
elliptical galaxy.   In the case of minor mergers (mass ratio $<$0.3), 
some works assign all the accreted gas to the disc 
while transferring all the accreted stars to the bulge (de Lucia et al.\ 2006), 
while others add both the stars and cold gas to the disc of the primary galaxy 
(Kauffmann \& Haehnelt 2000).  
The new bulge-to-disc flux ratio data may help
to refine how these semi-analytical models operate.

\subsubsection{Bulge profile shape}

Although equations to correct the S\'ersic parameters pertaining to the bulge
are not yet available, we note that the influence of dust at wavelengths
around 2 microns is noticeably less than in the optical bands 
(e.g. Whitford 1958; Cardelli, Clayton \& Mathis 1989; Fitzpatrick 1999). 
The significant change to $\mu_{0,K}$ with inclination which is seen in
Figure~\ref{fig-Dust} is not because of dust but predominantly due to the
greater line-of-sight depth through the (near-transparent) disc with
increasing disc inclination.  In the absence of dust, a spherical bulge will
look the same from all angles, i.e.\ inclinations.  We therefore do not expect
the $K$-band bulge parameters to change greatly (i.e.\ more than a few tenths
of a magnitude, or $\sim$20 per cent in size) with inclination,
and we report their observed values in Table~\ref{Tab_K}. 

As can be seen in Figure~\ref{figker}, Sc galaxies and earlier types typically
have values of $n \sim 2$ and sometimes as high as 4.  While Scd galaxies and
later types tend to have values of $n < 2$, some Sd galaxies can still have
values of n greater than 2.  It may therefore be misleading to label all Sc
and later-type galaxies as possessing pseudobulges built from exponential
discs (Erwin et al.\ 2003; Kormendy \& Kennicutt 2004, their section~4.2;
Athanassoula 2008).  While a value of $n=1$ in S\'ersic's model describes an
exponential profile, it does necessitate the presence of a flattened,
rotating, exponential disc rather than a spheroidal, pressure supported bulge.
Dwarf elliptical galaxies also have values of $n < 2$ (e.g.\ Caon et al.\
1993; Young \& Currie 1994; Jerjen et al.\ 2000) --- they define the low-mass
end of a linear relation between magnitude and S\'ersic index (e.g.\ 
Graham et al.\ 2006b, their figure~1; Nipoti,
Londrillo \& Ciotti 2006).  Dwarf ellipticals, however, are generally not
considered to have formed from the secular evolution of a disc.\footnote{Note:
  even if dwarf elliptical galaxies had formed from the violent redistribution
  of a disc, they need not preserve the disc's exponential radial profile
  shape.}  Subsequently, disc galaxy bulges with values of $n < 2 $ (or $n
\sim 1$) are not necessarily pseudobulges and should not be (re-)classified as
such based on this criteria alone.  To clarify Kormendy \& Kennicutt (2004),
while pseudobulges may have $n \sim 1$, a bulge with $n \sim 1$ need not be a
pseudobulge.
%

Although not a new result, the preponderance of bulges with $n<4$ combined
with the prevalence of late-type disc galaxies having bulges with $n \sim 1$ necessitates
that simulations of disc galaxies neither use nor create an $R^{1/4}$-like bulge 
for every galaxy. 
While the projection of Hernquist's (1990) useful model --- 
employed in many numerical simulations (e.g.\ Springel et al.\ 2005b; Di
Matteo et al.\ 2005) --- 
reproduces an $R^{1/4}$ profile, it does not provide a particularly good
representation of bulges, such as the Milky Way's bulge, which have $n\sim 1$ (e.g.\ 
Terzi\'c \& Graham 2005, their Figure~7; Terzi\'c \& Sprague 2007).
Furthermore, efforts to simulate a range of disc galaxies with different 
morphological types by only varying the bulge-to-disc mass ratio, but using a
Hernquist model for every bulge (e.g.\ Springel et al.\ 2005a), 
i.e.\ effectively assigning a de Vaucouleurs 
profile to every bulge, will not generate a realistic set of galaxies.
In the case of bulges formed from disc instabilities, one requires the
production of bulges that have exponential light profiles rather than
Hernquist-like density profiles.  In regard to simulations of minor mergers,
in which bulges are built up within discs, the observed correlation between
bulge profile shape and luminosity, and also supermassive black hole mass (Graham \& Driver 2007),
must be reproduced.  The recovery of bulge parameters using $R^{1/4}$-based
models is known to introduce systematic biases with profile shape and
thus also black hole mass (e.g.\ Trujillo et al.\ 2001, their Figures~4 and 5; Brown
et al.\ 2003, their Figure~7).  Therefore, once realistic bulges are
generated, the computation of their sizes and binding energy by assuming a
Hernquist model would be inappropriate and lead to spurious correlations when 
constructing, for example, a black hole fundamental plane (Younger et al.\ 2008).

Density models such as that from Einasto (1965) and Prugniel \& Simien
(1997) contain a ``shape parameter'' that effectively captures the range of
structural shapes (S\'ersic indices) which real bulges are observed to 
possess.\footnote{Kinematical expressions associated with these models,
  plus equations to convert projected densities such as $\langle \mu \rangle
  _{\rm e}$ into three-dimensional densities, are given in Terzi\'c \& Graham
  (2005), Merritt et al.\ (2006) and Graham et al.\ (2006a).}  
Such models have already provided insight into the gravitational
torques which oblate and triaxial bulges generate.  Interestingly, it is
bulges with smaller S\'ersic indices which have a greater non-axisymmetric
gravitational field: this is because bulges increasingly resemble central
point sources as the S\'ersic index increases (Trujillo et al.\ 2002).

By construction, 
the three-parameter ($\rho_e, R_{\rm e}, n$) density model of Prugniel \&
Simien (1997; their equation~B6) contains two identical parameters to 
S\'ersic's model and can be written as 
\begin{equation} \label{r1_app}
  \rho (r) = \rho_e \left({r\over R_{\rm e}}\right)^{-p}
             {\rm e}^{-b \left[ \left( r/R_{\rm e} \right)^{1/n} -1 \right] }, 
\end{equation}
\begin{equation} \label{r2_app}
  \rho_e = {M\over L} \ \frac{I_e}{R_{\rm e}} \ {b^{n(1-p)} \ 
           \Gamma(2n) \over 2 \Gamma(n(3-p)) },
\end{equation}
where $r$ is the internal, i.e.\ not projected, radius (see 
M\'arquez et al.\ 2001; Terzi\'c \&
Graham 2005, their equation~4).  For clarity, the subscript $n$ has been dropped from the 
term $b_n$.  The new parameter $\rho_e$ is the internal
density at $r=R_e$ and provides the normalisation such that the total mass
from equation~(\ref{r1_app}) equals that obtained from equation~(\ref{EqSB})
after applying the appropriate stellar mass-to-luminosity ratio $M/L$.
Equation~(\ref{r2_app}) has the same functional form as S\'ersic's model
although multiplied by an aditional power-law term with exponent $p=p(n)$.  We
adopt Lima Neto et al.'s (1999) expression for $p$, for which a 
high-quality match between deprojected S\'ersic profiles 
and the above expression is obtained when $p = 1.0 - 0.6097/n + 0.05563/n^2$.

After converting the observed $K$-band bulge effective radii into units of parsecs, 
and the observed $K$-band effective surface brightnesses into units of absolute solar 
luminosity per square parsec, we have used 
equation~(\ref{r2_app}) to derive the mean $\rho_{\rm e} / (M/L)$ values for
bulges of each 
galaxy type and class.  These luminosity densities are provided in
Table~\ref{Tab_K}. 
To do this we used an absolute $K$-band magnitude for the Sun of 3.33 mag (Cox 2000). 
Multiplying by one's preferred stellar $M/L$ ratio gives the stellar 
mass density of the bulge at $r=R_{\rm e}$. 
The mean $R_{\rm e}$, $n$ and $\rho_{\rm e}$ values can be used 
to construct realistic bulges using the Prugniel-Simien model, 
and/or to test if simulated bulges match the typical density profiles 
of real galaxy bulges.

\subsection{The optical}

As with the $K$-band data, we have collated literature data on the
structural parameters of galaxies measured at optical wavelengths
(see Table~\ref{TabLit}).  After first removing any inclination corrections
applied in these papers, we again employed equations~\ref{Eq_mu}
and \ref{Eq_h} to determine the disc central surface brightnesses 
and scalelengths, and we computed the disc magnitudes using equation~\ref{Eq_disc}.
The results are shown in Tables~\ref{Tab_B} to \ref{Tab_I}.

\begin{table*}
\caption{
$B$ band parameters. 
The median, $\pm$68/2 per cent of the distribution on either side of the median, is
shown as a function of galaxy type (Column 1).
Columns 2 and 3: Face-on, disc central surface brightness and scalelength corrected
using equations~\ref{Eq_mu} and \ref{Eq_h}, respectively.
Column 4: Disc magnitude computed using equation~\ref{Eq_disc} and the entries in columns 2 and 3.
Column 5: Bulge magnitude, obtained from the observed flux, corrected using equation~\ref{Eq_Mb}. 
Column 6: Disc magnitude, obtained from the observed flux, corrected using equation~\ref{Eq_Md}. 
Column 7: Bulge-to-disc flux ratio using columns 5 and 6.
Column 8: Number of data points. 
\label{Tab_B}
}
\begin{tabular}{lccccccc}
\hline
Type &  $\mu_0$ & $h$ & $M_{\rm disc}$ &  $M_{\rm bulge}$ & $M_{\rm disc}$ &  $\log(B/D)$ &  \# \\
1 &  2  & 3   & 4  & 5  &  6 &  7  & 8  \\
\hline
\multicolumn{8}{c}{Morphological Type} \\
$-3,-2,-1$,S0    &            ...          &            ...         &  ...               &             ...            &            ...            &  ...           &  0 \\   
 0, S0/a         & $21.48^{+0.32}_{-0.47}$ & $6.54^{+3.47}_{-2.49}$ & $-21.95^{+1.72}_{-0.15}$ &   $-20.81^{+1.16}_{-0.41}$ &  $-22.34^{+1.81}_{-0.16}$ &  $-0.46^{+0.16}_{-0.10}$ &  3 \\  
 1, Sa           & $20.95^{+1.18}_{-0.38}$ & $3.00^{+6.70}_{-0.85}$ & $-20.21^{+0.94}_{-1.40}$ &   $-20.08^{+2.06}_{-0.99}$ &  $-20.57^{+0.76}_{-1.56}$ &  $-0.29^{+0.43}_{-0.41}$ &  9 \\  
 2, Sab          & $20.89^{+0.95}_{-0.64}$ & $5.05^{+5.10}_{-2.57}$ & $-20.93^{+0.83}_{-0.97}$ &   $-20.38^{+1.22}_{-1.23}$ &  $-21.24^{+0.75}_{-1.20}$ &  $-0.39^{+0.27}_{-0.51}$ & 10 \\  
 3, Sb           & $20.77^{+0.81}_{-0.56}$ & $3.68^{+4.29}_{-1.77}$ & $-21.05^{+1.45}_{-0.80}$ &   $-19.50^{+2.16}_{-1.02}$ &  $-21.65^{+1.51}_{-0.84}$ &  $-0.87^{+0.44}_{-0.42}$ & 32 \\  
 4, Sbc          & $20.72^{+0.57}_{-0.66}$ & $3.86^{+2.88}_{-1.98}$ & $-20.71^{+1.14}_{-0.87}$ &   $-17.76^{+1.40}_{-2.26}$ &  $-21.11^{+1.00}_{-0.91}$ &  $-1.13^{+0.42}_{-0.51}$ & 40 \\  
 5, Sc           & $20.60^{+0.65}_{-0.70}$ & $3.37^{+1.66}_{-0.88}$ & $-20.75^{+0.66}_{-0.46}$ &   $-18.04^{+1.32}_{-1.11}$ &  $-21.14^{+0.68}_{-0.50}$ &  $-1.28^{+0.46}_{-0.24}$ & 49 \\  
 6, Scd          & $21.23^{+0.35}_{-1.26}$ & $3.36^{+1.17}_{-1.56}$ & $-20.27^{+1.40}_{-0.43}$ &   $-16.41^{+0.63}_{-1.47}$ &  $-20.63^{+1.43}_{-0.44}$ &  $-1.55^{+0.24}_{-0.30}$ & 16 \\  
 7, Sd           & $22.10^{+0.86}_{-0.49}$ & $3.32^{+1.80}_{-0.83}$ & $-19.27^{+0.92}_{-1.21}$ &   $-16.24^{+1.70}_{-1.02}$ &  $-19.53^{+0.80}_{-1.29}$ &  $-1.55^{+0.56}_{-0.39}$ & 13 \\  
 8, Sdm          & $21.74^{+0.76}_{-0.22}$ & $3.88^{+1.08}_{-1.58}$ & $-19.68^{+1.43}_{-0.30}$ &   $-15.88^{+1.36}_{-0.66}$ &  $-20.11^{+1.41}_{-0.27}$ &  $-1.69^{+0.53}_{-0.24}$ &  6 \\  
 9, Sm           & $22.61^{+0.75}_{-0.46}$ & $3.72^{+0.71}_{-1.37}$ & $-18.42^{+0.66}_{-0.88}$ &   $-15.33^{+1.23}_{-0.37}$ &  $-18.69^{+0.51}_{-1.10}$ &  $-1.53^{+0.16}_{-0.14}$ &  7 \\  
10, Irr          & $22.25^{+0.00}_{-0.00}$ & $2.06^{+0.00}_{-0.00}$ & $-17.91^{+0.00}_{-0.00}$ &   $-14.27^{+0.00}_{-0.00}$ &  $-18.18^{+0.00}_{-0.00}$ &  $-1.57^{+0.00}_{-0.00}$ &  1 \\  
\multicolumn{8}{c}{Morphological Class} \\
$-3\leq T\leq 0$ & $21.48^{+0.32}_{-0.47}$ & $6.54^{+3.47}_{-2.49}$ & $-21.95^{+1.72}_{-0.15}$ &   $-20.81^{+1.16}_{-0.41}$ &  $-22.34^{+1.81}_{-0.16}$ &  $-0.46^{+0.16}_{-0.10}$ & 3 \\   
$T=1$-3          & $20.80^{+1.13}_{-0.57}$ & $3.76^{+5.69}_{-1.84}$ & $-20.87^{+1.32}_{-0.97}$ &   $-19.61^{+1.76}_{-1.26}$ &  $-21.34^{+1.31}_{-1.07}$ &  $-0.71^{+0.59}_{-0.50}$ & 51 \\  
$T=6$-9          & $21.77^{+0.96}_{-1.12}$ & $3.67^{+1.18}_{-1.71}$ & $-19.47^{+1.23}_{-1.18}$ &   $-15.93^{+1.42}_{-1.55}$ &  $-19.83^{+1.22}_{-1.14}$ &  $-1.58^{+0.38}_{-0.32}$ & 42 \\  
\hline
\end{tabular}
\end{table*}

\begin{table}
\caption{
$V$ band parameters. 
The median, $\pm$68/2 per cent of the distribution on either side of the median, is
shown as a function of galaxy type (Column 1).
Columns 2 and 3: Face-on, disc central surface brightness and scalelength corrected
using equations~\ref{Eq_mu} and \ref{Eq_h}, respectively.
Column 4: Disc magnitude computed using equation~\ref{Eq_disc} and the entries in column 2 and 3.
Column 5: Number of data points. 
\label{Tab_V}
}
\begin{tabular}{lcccc}
\hline
Type &  $\mu_0$ & $h$ & $M_{\rm disc}$  &  \# \\
1 &  2  & 3   & 4  & 5   \\
\hline
\multicolumn{5}{c}{Morphological Type} \\
$-3,-2,-1$,S0    &          ...            &          ...           &          ...             & 0 \\
 0, S0/a         & $21.08^{+0.00}_{-0.00}$ & $9.91^{+0.00}_{-0.00}$ & $-22.70^{+0.00}_{-0.00}$ & 1 \\
 1, Sa           & $20.15^{+1.57}_{-0.06}$ & $3.15^{+5.75}_{-0.62}$ & $-20.94^{+0.82}_{-1.16}$ & 7 \\
 2, Sab          & $20.44^{+0.41}_{-1.11}$ & $4.12^{+2.90}_{-1.76}$ & $-21.29^{+0.62}_{-0.68}$ & 5 \\
 3, Sb           & $19.90^{+0.57}_{-0.43}$ & $4.20^{+2.21}_{-1.35}$ & $-21.50^{+0.49}_{-0.88}$ & 19 \\
 4, Sbc          & $20.04^{+0.47}_{-1.06}$ & $3.78^{+2.29}_{-2.26}$ & $-21.26^{+0.80}_{-0.55}$ & 15 \\
 5, Sc           & $19.78^{+0.56}_{-0.19}$ & $3.10^{+1.05}_{-0.48}$ & $-21.45^{+0.96}_{-0.26}$ & 20 \\
 6, Scd          & $20.03^{+0.58}_{-0.80}$ & $3.24^{+0.44}_{-2.01}$ & $-20.72^{+0.98}_{-0.51}$ & 9 \\
 7, Sd           & $21.75^{+0.14}_{-0.89}$ & $3.58^{+1.24}_{-0.65}$ & $-19.52^{+0.24}_{-1.60}$ & 5 \\
 8, Sdm          & $21.13^{+0.25}_{-0.25}$ & $2.74^{+0.81}_{-0.81}$ & $-21.14^{+0.70}_{-0.70}$ & 2 \\
 9, Sm           &                         &                        &                          & 0 \\
10, Irr          &                         &                        &                          & 0 \\
\multicolumn{5}{c}{Morphological Class} \\ 
$-3\leq T\leq 0$ & $21.08^{+0.00}_{-0.00}$ & $9.91^{+0.00}_{-0.00}$ & $-22.70^{+0.00}_{-0.00}$ & 1 \\ 
$T=1$-3          & $20.09^{+0.93}_{-0.63}$ & $3.83^{+3.90}_{-1.36}$ & $-21.49^{+0.76}_{-0.86}$ & 31 \\
$T=6$-9          & $20.59^{+1.06}_{-1.34}$ & $3.34^{+0.97}_{-1.69}$ & $-20.68^{+1.25}_{-1.14}$ & 16 \\
\hline
\end{tabular}
\end{table}

\begin{table}
\caption{
$R$ band parameters. 
The median, $\pm$68/2 per cent of the distribution on either side of the median, is
shown as a function of galaxy type (Column 1).
Column 2: Bulge magnitude, obtained from the observed flux corrected using equation~\ref{Eq_Mb}. 
Column 3: Disc magnitude, obtained from the observed flux corrected using equation~\ref{Eq_Md}. 
Column 4: Bulge-to-disc flux ratio using columns 2 and 3.
Column 5: Number of data points. 
\label{Tab_R}}
\begin{tabular}{lcccc}
\hline
Type &  $M_{\rm bulge}$ & $M_{\rm disc}$ &  $\log(B/D)$ &  \# \\
1 &  2  & 3   & 4  & 5   \\
\hline
\multicolumn{5}{c}{Morphological Type} \\
$-3,-2,-1$,S0    &   ...                     &    ...                   &     ...                   &    0 \\
 0, S0/a         &  $-22.45^{+1.35}_{-0.41}$ &  $-23.51^{+1.50}_{-0.03}$ &  $-0.34^{+0.10}_{-0.07}$ &    3 \\
 1, Sa           &  $-21.96^{+2.15}_{-0.45}$ &  $-21.96^{+0.88}_{-0.86}$ &  $-0.28^{+0.47}_{-0.08}$ &   10 \\
 2, Sab          &  $-21.28^{+1.83}_{-1.39}$ &  $-21.99^{+0.64}_{-1.43}$ &  $-0.33^{+0.48}_{-0.89}$ &    9 \\
 3, Sb           &  $-20.58^{+2.50}_{-1.32}$ &  $-22.65^{+1.36}_{-0.59}$ &  $-0.72^{+0.35}_{-0.34}$ &   35 \\
 4, Sbc          &  $-19.48^{+1.19}_{-1.89}$ &  $-22.27^{+1.19}_{-0.76}$ &  $-0.96^{+0.42}_{-0.44}$ &   36 \\
 5, Sc           &  $-19.33^{+1.13}_{-1.27}$ &  $-22.13^{+0.90}_{-0.53}$ &  $-1.07^{+0.45}_{-0.30}$ &   47 \\
 6, Scd          &  $-18.21^{+1.27}_{-0.98}$ &  $-21.48^{+1.47}_{-0.48}$ &  $-1.38^{+0.36}_{-0.12}$ &   19 \\
 7, Sd           &  $-17.22^{+1.22}_{-1.39}$ &  $-20.28^{+0.72}_{-1.63}$ &  $-1.38^{+0.47}_{-0.50}$ &   18 \\
 8, Sdm          &  $-16.75^{+0.83}_{-1.21}$ &  $-20.61^{+0.75}_{-0.44}$ &  $-1.46^{+0.21}_{-0.19}$ &    7 \\
 9, Sm           &  $-16.61^{+1.05}_{-0.59}$ &  $-19.50^{+0.60}_{-1.19}$ &  $-1.16^{+0.03}_{-0.27}$ &    7 \\
10, Irr          &  $-16.77^{+0.68}_{-0.68}$ &  $-19.23^{+0.35}_{-0.35}$ &  $-0.99^{+0.13}_{-0.13}$ &    2 \\
						                       	    	             	   	  
$-3\leq T\leq 0$ &  $-22.45^{+1.35}_{-0.41}$ &  $-23.51^{+1.50}_{-0.03}$ &  $-0.34^{+0.10}_{-0.07}$ &    3 \\
$T=1$-3          &  $-21.04^{+1.94}_{-1.14}$ &  $-22.44^{+1.23}_{-0.86}$ &  $-0.60^{+0.44}_{-0.42}$ &   54 \\
$T=6$-9          &  $-17.44^{+1.50}_{-1.21}$ &  $-20.61^{+1.07}_{-1.28}$ &  $-1.36^{+0.32}_{-0.33}$ &   51 \\
\multicolumn{5}{c}{Morphological Class} \\
\hline
\end{tabular}
\end{table}

\begin{table*}
\caption{
$I$ band parameters. 
The median, $\pm$68/2 per cent of the distribution on either side of the median, is
shown as a function of galaxy type (Column 1).
Columns 2 and 3: Face-on, disc central surface brightness and scalelength corrected
using equations~\ref{Eq_mu} and \ref{Eq_h}, respectively.
Column 4: Disc magnitude computed using equation~\ref{Eq_disc} and the entries in columns 2 and 3.
Column 5: Bulge magnitude, obtained from the observed flux, corrected using equation~\ref{Eq_Mb}. 
Column 6: Disc magnitude, obtained from the observed flux, corrected using equation~\ref{Eq_Md}. 
Column 7: Bulge-to-disc flux ratio using columns 5 and 6.
Column 8: Number of data points. 
\label{Tab_I}
}
\begin{tabular}{lccccccc}
\hline
Type &  $\mu_0$ & $h$ & $M_{\rm disc}$ &  $M_{\rm bulge}$ & $M_{\rm disc}$ &  $\log(B/D)$ &  \# \\
1 &  2  & 3   & 4  & 5  &  6 &  7  & 8  \\
\hline
\multicolumn{8}{c}{Morphological Type} \\
$-3,-2,-1$,S0    &            ...           &            ...          &  ...                       &  ...                      &            ...            &  ...                    &   0 \\   
 0, S0/a         &  $20.17^{+0.37}_{-0.37}$ &  $7.63^{+2.84}_{-2.84}$ &  $-22.56^{+0.60}_{-0.60}$  & $-22.31^{+0.89}_{-0.89}$  & $-22.84^{+0.65}_{-0.65}$  & $-0.21^{+0.09}_{-0.09}$ &   2 \\ 
 1, Sa           &  $19.49^{+0.89}_{-1.03}$ &  $3.43^{+5.07}_{-1.35}$ &  $-21.92^{+0.77}_{-1.09}$  & $-22.49^{+2.40}_{-0.58}$  & $-22.41^{+0.83}_{-0.98}$  & $-0.24^{+0.42}_{-0.30}$ &   9 \\ 
 2, Sab          &  $19.42^{+0.65}_{-0.80}$ &  $5.32^{+1.97}_{-3.32}$ &  $-22.41^{+1.02}_{-0.88}$  & $-22.15^{+1.04}_{-0.93}$  & $-22.66^{+0.99}_{-0.98}$  & $-0.16^{+0.26}_{-0.68}$ &   8 \\ 
 3, Sb           &  $19.07^{+0.24}_{-0.90}$ &  $3.85^{+3.02}_{-1.98}$ &  $-22.77^{+1.51}_{-0.51}$  & $-21.62^{+1.64}_{-1.00}$  & $-23.26^{+1.76}_{-0.43}$  & $-0.53^{+0.27}_{-0.30}$ &  16 \\ 
 4, Sbc          &  $19.28^{+0.40}_{-1.10}$ &  $4.41^{+2.73}_{-2.95}$ &  $-22.41^{+1.20}_{-1.01}$  & $-20.74^{+1.84}_{-1.34}$  & $-22.79^{+1.30}_{-0.83}$  & $-0.86^{+0.34}_{-0.40}$ &  25 \\ 
 5, Sc           &  $19.03^{+0.83}_{-0.46}$ &  $3.40^{+1.08}_{-0.92}$ &  $-22.32^{+1.10}_{-0.49}$  & $-19.98^{+1.34}_{-1.54}$  & $-22.59^{+1.24}_{-0.52}$  & $-1.06^{+0.62}_{-0.28}$ &  30 \\ 
 6, Scd          &  $19.70^{+0.74}_{-0.08}$ &  $2.93^{+0.71}_{-1.13}$ &  $-21.24^{+1.45}_{-0.42}$  & $-17.38^{+0.56}_{-1.74}$  & $-21.51^{+1.14}_{-0.69}$  & $-1.34^{+0.22}_{-0.11}$ &   7 \\ 
 7, Sd           &  $21.05^{+0.24}_{-1.77}$ &  $3.66^{+1.72}_{-0.67}$ &  $-20.90^{+0.83}_{-1.17}$  & $-18.50^{+1.38}_{-0.48}$  & $-21.11^{+0.87}_{-1.27}$  & $-1.35^{+0.51}_{-0.10}$ &   9 \\ 
 8, Sdm          &  $20.18^{+0.54}_{-0.29}$ &  $4.00^{+0.26}_{-1.51}$ &  $-20.92^{+0.95}_{-0.57}$  & $-17.48^{+1.69}_{-0.89}$  & $-21.10^{+0.94}_{-0.75}$  & $-1.68^{+0.41}_{-0.16}$ &   4 \\ 
 9, Sm           &  $21.10^{+0.44}_{-0.08}$ &  $4.23^{+0.51}_{-1.97}$ &  $-20.36^{+1.51}_{-0.29}$  & $-17.77^{+2.75}_{-0.54}$  & $-20.70^{+1.11}_{-0.42}$  & $-1.17^{+0.05}_{-0.65}$ &   3 \\ 
10, Irr          &  $21.14^{+0.00}_{-0.00}$ &  $1.99^{+0.00}_{-0.00}$ &  $-18.94^{+0.00}_{-0.00}$  & $-15.73^{+0.00}_{-0.00}$  & $-19.16^{+0.00}_{-0.00}$  & $-1.37^{+0.00}_{-0.00}$ &   1 \\ 
\multicolumn{8}{c}{Morphological Class} \\ 
$-3\leq T\leq 0$ &  $20.17^{+0.37}_{-0.37}$ &  $7.63^{+2.84}_{-2.84}$ &  $-22.56^{+0.60}_{-0.60}$  & $-22.31^{+0.89}_{-0.89}$  & $-22.84^{+0.65}_{-0.65}$  & $-0.21^{+0.09}_{-0.09}$  &  2 \\ 
$T=1$-3          &  $19.11^{+0.97}_{-0.67}$ &  $3.91^{+3.42}_{-2.07}$ &  $-22.55^{+1.37}_{-0.78}$  & $-21.78^{+1.88}_{-1.05}$  & $-22.91^{+1.47}_{-0.77}$  & $-0.39^{+0.43}_{-0.38}$  & 33 \\ 
$T=6$-9          &  $20.43^{+0.77}_{-0.77}$ &  $3.62^{+1.20}_{-1.50}$ &  $-20.79^{+1.01}_{-1.12}$  & $-17.92^{+1.32}_{-1.04}$  & $-21.11^{+1.03}_{-1.19}$  & $-1.35^{+0.29}_{-0.46}$  & 23 \\ 
\hline
\end{tabular}
\end{table*}

\begin{figure}
\includegraphics[angle=270,scale=0.41]{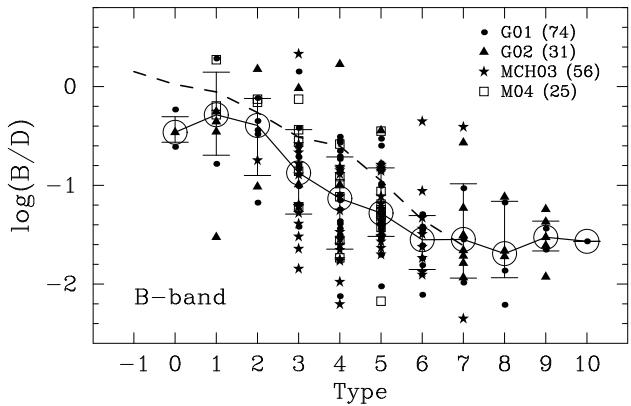}
\caption{ 
Logarithm of the $B$-band, bulge-to-disc flux ratio 
as a function of galaxy Type (see Table~\ref{Tab_B}, column~7). 
The dashed line traces the $B$-band values from Simien \& de Vaucouleurs
(1986). 
}
\label{fig-B-BD}
\end{figure}

While dust is expected to modify the individual bulge S\'ersic parameters at
optical wavelengths --- and for this reason we don't present such values here
--- we can apply the magnitude corrections from equation~\ref{Eq_Mb}
to the observed bulge magnitudes.  Table~\ref{Tab_B} shows these $B$-band
bulge (and disc) magnitudes and the logarithm of the ratio of their
luminosities.  This latter quantity is plotted in Figure~\ref{fig-B-BD}. 
The $B$-band flux ratios obtained from Simien \& de Vaucouleurs (1986), which are still used
by many research groups (e.g.\ Springel et al.\ 2001; 
Mathis et al.\ 2002; De Lucia et al.\ 2006) to build galaxies and assign
galaxy types, are up to 0.5 dex larger. 
If the $R^{1/4}$-derived flux ratios from Simien \& Vaucouleurs were corrected 
for dust, then they would be greater still.  While our galaxy parameters have
been corrected for dust, they have also been derived using $R^{1/n}$
rather than $R^{1/4}$ bulge models which, collectively, results in smaller
flux ratios.

\begin{figure}
\includegraphics[angle=270,scale=0.46]{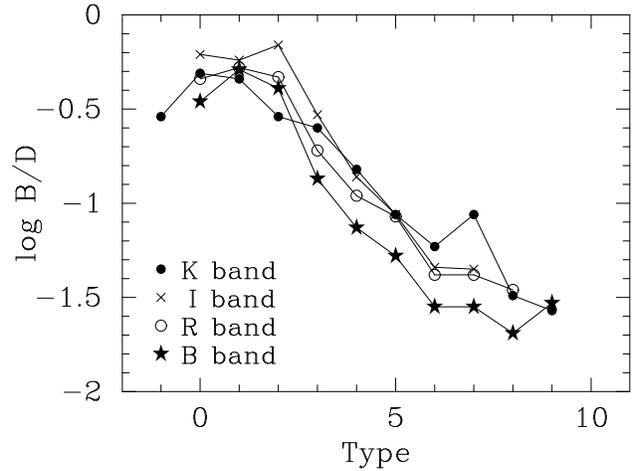}
\caption{
Median logarithm of the bulge-to-disc flux ratio as a function of 
galaxy Type. 
The number of galaxies involved in each data point can be seen in 
Tables~\ref{Tab_Ratio} to \ref{Tab_I}. 
}
\label{fig-B-K-BD}
\end{figure}

Having corrected the $B$-band $B/D$ ratios for dust does not imply that they
should be equal to the $K$-band values.  In fact, differences are expected and
will reflect the varying stellar populations of the bulges and discs.  For
example, a redder intrinsic (i.e. before the effects of dust) SED for a bulge
will result in a larger $B/D$ ratio in the $K$-band than in the $B$-band.  In
Figure~\ref{fig-B-K-BD} we have co-plotted the $K$- and $B$-band $B/D$ values
as a function of galaxy type.  The S0--Sab galaxies have roughly the same
$B/D$ ratio, possibly indicating similar stellar population, i.e.\ if the
bulge is red, then the disc will also be red (Peletier \& Balcells 1996).
Although we do caution that other factors may be at play.  Laurikainen et al.\
(2005) were careful to separate the bar flux from the bulge flux in their
$K$-band analysis of early-type disc galaxies while Balcells et al.'s (2003)
$K$-band sample did not possess bars.  The optical studies might have overly
large $B/D$ ratios due to the assignment of bar flux to the bulge.

The later type spiral galaxies tend to have a $B$-band $\log(B/D)$ value which
is nearly some $\sim$0.3 dex smaller than the $K$-band value.  That is, for
the intermediate and late-type spiral galaxies, the $B$-band $B/D$ ratio is a
factor of 2 smaller than the $K$-band ratio.  The obvious interpretation is
that the later types have bluer --- younger and/or more metal poor --- discs
relative to their bulges.
Regarding the Sm and irregular galaxies, the situation becomes somewhat mixed,
and the diagram also suffers from low number statistics at this end.
To provide a rough idea of the uncertainty on the plotted data points in
Figure~\ref{fig-B-K-BD}, we note that for a distribution with sample standard
deviation equal to $\sigma$, the uncertainty on the mean is $\sigma/\sqrt(N)$,
where $N$ is the number of data points.  From Table~\ref{Tab_Ratio} we have
$\sigma \sim 0.4$, while the value of $N$ can be seen in
Tables~\ref{Tab_Ratio} to \ref{Tab_I}.

%
%

\section{Bivariate distributions} \label{Sec_Double}

Scaling relations, if found to exist, can {\it (i)} provide insight into the
physical mechanisms which shape galaxies and {\it (ii)} supply helpful
restrictions for simulations which try to match the real Universe and thereby
understand it.  In this section we present dust-corrected brightness-size
distributions for galaxy discs (and also bulges and elliptical galaxies).
Because the majority of disc galaxies consist of two main, physically
distinct, components --- a three-dimensional bulge and a relatively flat,
two-dimensional disc --- we do not treat the disc galaxies as single entities.
Instead, we separately show bivariate distributions for the structural
properties of the bulges and the discs.

The data from Dong \& De Robertis (2006) is not included from here on.  Their
challengingly-shallow (eight second exposure) 2MASS images may be responsible
for a number of apparent outliers which would slightly bias the fits in the
following section if they had been included.

\subsection{The $\mu_0$-$\log h$ distribution}\label{SecMuh}

\begin{figure}
\includegraphics[angle=0,scale=0.54]{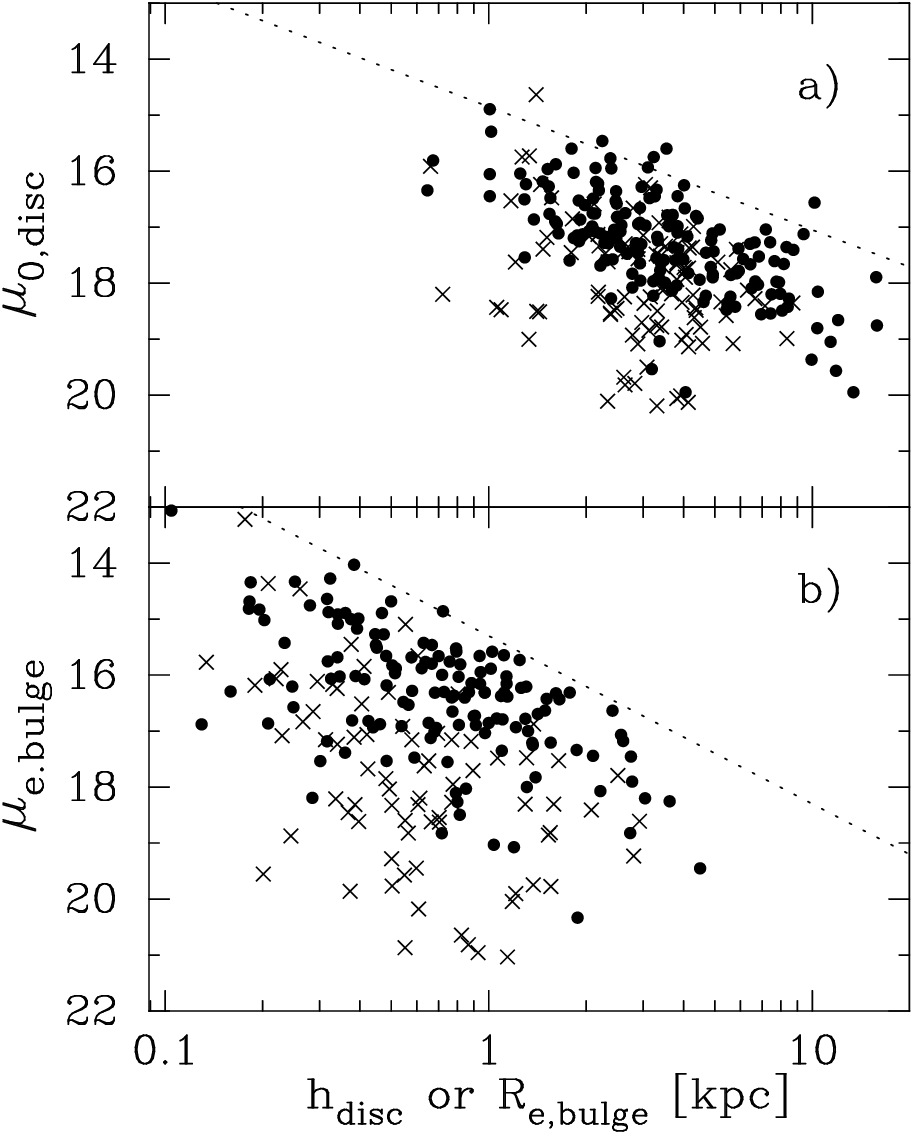}
\caption{ $K$-band. 
Top panel: Central disc surface brightness (corrected using equation~\ref{Eq_mu})
versus disc scalelength $h$ (corrected using equation~\ref{Eq_h}).
Bottom panel: Bulge effective surface brightness (observed) 
versus the bulge effective radius $R_{\rm e}$ (observed). 
The upper bright envelope is traced here with the (empirical) lines 
$\mu_{\rm 0,bright} = 14.85 + 2.2 \log h$ (panel a) and 
$\mu_{\rm e,bright} = 15.3 + 3.0 \log R_{\rm e}$ (panel b).
Galaxy types Sbc ($T=4$) and earlier are
denoted by the circles, while later galaxy types are denoted by the crosses.
}
\label{fig-muer}
\end{figure}

Figure~\ref{fig-muer}a shows the $K$-band distribution of face-on, 
central disc surface brightness $\mu_0$ versus disc scalelength $h$.  
Both of these parameters have been fully corrected for the influence of dust. 
For exponential discs, one can easily transform this diagram so that it
involves the effective radii, $R_{\rm e}$, and effective surface brightnesses, 
$\mu_e$.  This is achieved 
by using $R_{\rm e,disc} = 1.678 h$ and $\mu_{\rm e.disc} = \mu_0 + 1.822$, or
the mean effective surface brightness within $R_{\rm e,disc}$, which is such
that $\langle \mu \rangle _{\rm e,disc} = \mu_0 + 1.123$.

\begin{figure}
\includegraphics[angle=270,scale=0.39]{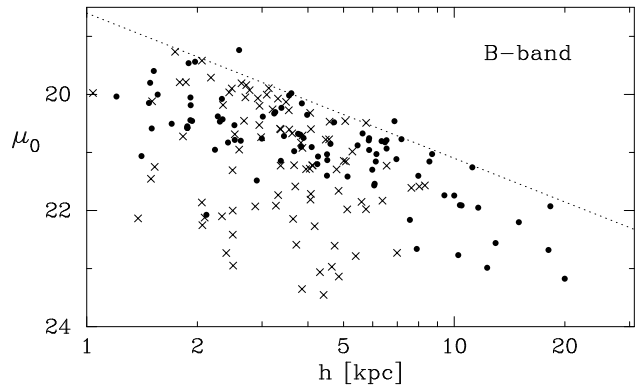}
\caption{  $B$-band. 
Central disc surface brightness (corrected using equation~\ref{Eq_mu})
versus disc scalelength (corrected using equation~\ref{Eq_h}). 
Rather than a Freeman (1970) law of constant surface brightness, an upper
bright envelope, which has been seen before (e.g.\ Graham 2001b and references
therein), is evident. 
The empirically-fitted dotted line in this diagram is such that 
$\mu_{0,bright} = 18.6 + 2.5 \log h$. 
Galaxy types Sbc ($T=4$) and earlier are 
denoted by circles, while later galaxy types are denoted by crosses.
The noticeably broader distribution for the late-type disc galaxies has also 
been noted before (Graham \& de Blok 2001). 
}
\label{fig-murky}
\end{figure}

Due to selection effects, small faint galaxies are likely to be missed and
thus the lower-left of Figure~\ref{fig-muer}a may be missing galaxies. 
The upper envelope in Figure~\ref{fig-muer}a
does however reflect a real cutoff in the distribution of galactic
parameters, simply because bigger and brighter discs would be seen if they existed. 
The late-type disc galaxies can be seen to occupy a much broader region of this diagram than
the early-type disc galaxies, and as such they are not particularly well
represented by a linear surface brightness-size relation.
Moreover, given the differing distribution of the early- and late-type disc galaxies 
(see also Figure~\ref{fig-murky}), one can see that it is predominantly the
late-type disc galaxies for which we may have an incomplete distribution
in the lower-left quadrant.

Figure~\ref{fig-murky} shows the $B$-band $\mu_0$--$\log h$ relation in which
the same general behaviour is observed.  One of the immediate results that can 
be drawn from these figures is that the Freeman (1970) law, pertaining to a
constant central disc surface brightness, requires modification.  Not only do 
low surface brightness galaxies exist with fainter surface brightnesses ---
which has long been known (e.g.\ Disney 1976, 1998) --- but the bright limit
does not simply have the canonical Freeman surface brightness of 21.65 B-mag
arcsec$^{-2}$.  Instead, the bright limit is observed to depend strongly on 
disc size such that $\mu_{\rm 0,bright} \sim 2.5 \log h$
(Graham 2001b; Graham \& de Blok 2001, and references therein). 
Obviously, this relation (quantified in section~\ref{Sec_LR}) also implies an
improvement over the simple use of average disc scalelengths and central
surface brightnesses to represent each of the (early-type) disc galaxy types.

\subsection{The $\mu_{\rm e}$-$\log R_{\rm e}$ distribution}

The $K$-band distribution of bulge effective surface brightnesses $\mu_{\rm
e}$ and effective radii $R_{\rm e}$ are plotted in Figure~\ref{fig-muer}b.  As
with the disc parameters, the bright envelope to this distribution is real and
not an artifact of galaxy selection criteria.  The
bright boundary to the distribution is roughly described by the empirical
relation $\mu_{\rm e,bright} \sim 3 \log R_{\rm e}$.
Figure~\ref{fig-muer}b qualitatively agrees with
the (surface brightness)- size diagram in Binney \& Merrifield (1998, their
Fig.4.52) which was based on optical images of 66 galaxies from Kent (1985).
The advances here are (i) sample size (ii) the influence of dust is less in
the $K$-band and (iii) $R^{1/n}$ rather than $R^{1/4}$ models have been
applied to every bulge.

\subsection{The luminosity-size distribution}
\label{Sec_LR}

\begin{figure}
\includegraphics[angle=0,scale=0.85]{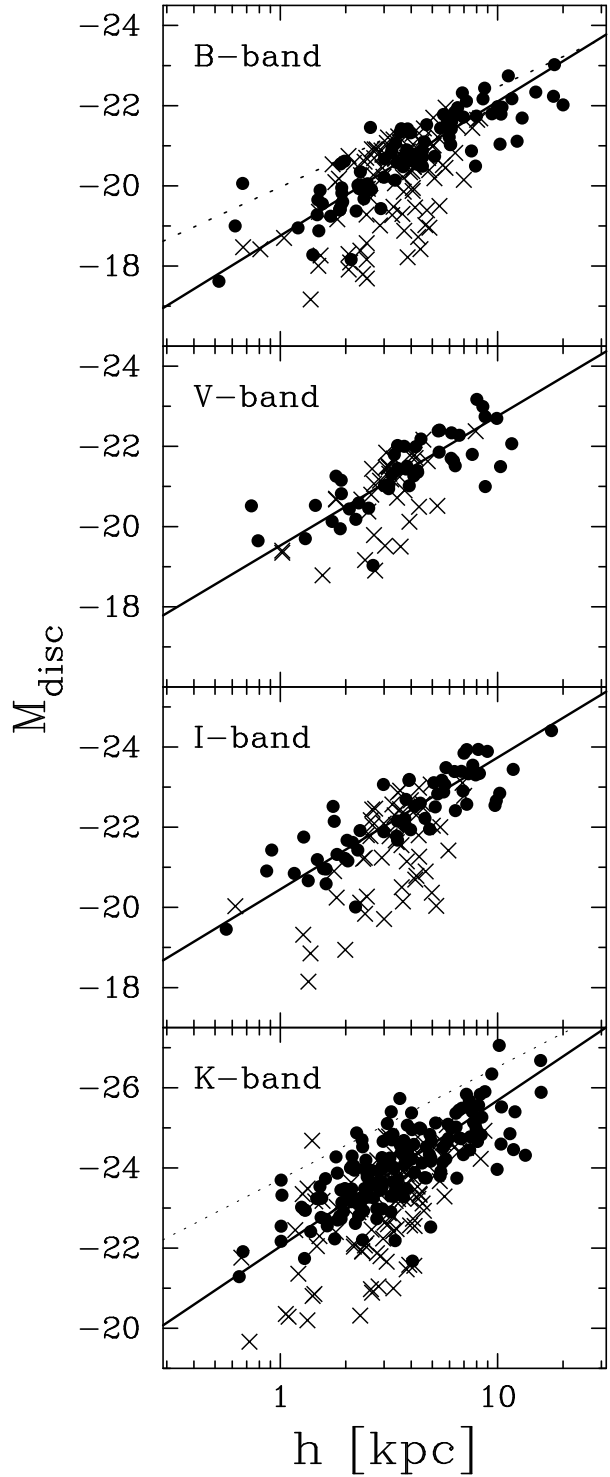}
\caption{ 
Inclination- and dust-corrected magnitude 
versus disc scalelength (via equation~\ref{Eq_h}).  
Galaxy types Sbc ($T=4$) and earlier are
denoted by the circles, while later galaxy types are denoted by the crosses. 
The differing distributions between the galaxy types in these figures 
is directly related to their 
different distributions in the $\mu_0$-$\log h$ plane (Figures~\ref{fig-muer}
and \ref{fig-murky}). 
Similarly, evidence of a relation here is simply a re-representation of 
evidence for a relation in the $\mu_0$-$\log h$ plane. 
The fitted lines to the discs of the early-type disc galaxies are given by 
the bisector regression of equation~\ref{Eq_LR} (see Table~\ref{Tab_LR}).  
The dotted lines seen here denote the bright boundaries shown in Figures~\ref{fig-muer}a 
and \ref{fig-murky}. 
}
\label{Fig_LR}
\end{figure}

Given that $L=2\pi I_0 h^2$ for exponential discs, the size-luminosity diagram is 
simply an alternative representation of the more traditionally used (surface
brightness)-size diagrams seen in section~\ref{SecMuh}.  For convenience, both
are shown here, although we emphasize that the upcoming size-luminosity relations could have
been equally as well obtained by fitting lines to the data in Figures~\ref{fig-muer}a and
\ref{fig-murky}.  The dashed lines in those figures have been mapped
into Figure~\ref{Fig_LR} which shows the (dust-corrected) disc magnitudes 
versus the (dust-corrected) disc scalelengths.

The early--type disc galaxy distribution in Figure~\ref{Fig_LR}
has been fit with a linear relation such that 
\begin{equation}
\log h = a - \frac{b}{2.5}(M_{\rm disc}+M_0), 
\label{Eq_LR}
\end{equation}
and thus $h \propto L^b$.  The quantity $M_0$ is simply a constant to help 
centralise the fit and thereby reduce the uncertainty on the intercept $a$.
We have employed Feigelson \& Babu's (1992) code SLOPES to perform six
regression analyses: 
ordinary least squares of Y on X, OLS(Y$\mid$X); 
and the inverse OLS(X$\mid$Y); 
the line which bisects these two; 
minimisation of the perpendicular residuals (known as orthogonal regression);
the geometric mean of OLS(Y$\mid$X) and OLS(X$\mid$Y), referred to as the 
  reduced major axis regression line; 
and finally the arithmetic mean of the two OLS lines. 
Here, $Y = \log h$ and $X = M_{\rm disc}$, and the results of the regression
are shown in Tables~\ref{Tab_LR} and \ref{Tab_KLR}.  
Table~\ref{Tab_KLR} also shows the 
results of the regressions for different galaxy types, 
although only in the $K$-band due to the numbers of galaxies involved.

\begin{table*}
\caption{
Linear regression coefficients for the size-luminosity relations
of disc galaxies with  $T \leq 4$ (i.e., $\leq$Sbc).  The fitted 
relation is such that $\log h = a - 0.4b(M_{\rm disc}+M_0)$. 
The OLS bisector lines of regression are shown in Figure~\ref{Fig_LR}.
\label{Tab_LR}
}
\begin{tabular}{lcc cc cc cc}
\hline
Method & \multicolumn{2}{c}{$B$-band ($M_0=20$)} & \multicolumn{2}{c}{$V$-band ($M_0=21$)} & \multicolumn{2}{c}{$I$-band ($M_0=22$)} & \multicolumn{2}{c}{$K$-band ($M_0=24$)} \\
                   &       $a$      &      $b$       &         $a$    &       $b$      &         $a$    &      $b$       &       $a$      &     $b$   \\
\hline
OLS(Y$\mid $X)     &  $0.40\pm0.02$ & $0.65\pm0.04$  &  $0.48\pm0.03$ & $0.59\pm0.08$  &  $0.49\pm0.02$ & $0.65\pm0.05$  &  $0.54\pm0.01$ & $0.49\pm0.04$ \\
OLS(X$\mid $Y)     &  $0.34\pm0.03$ & $0.86\pm0.06$  &  $0.42\pm0.04$ & $1.00\pm0.14$  &  $0.46\pm0.03$ & $0.89\pm0.07$  &  $0.54\pm0.02$ & $0.93\pm0.07$ \\
OLS bisector       &  $0.37\pm0.02$ & $0.75\pm0.05$  &  $0.46\pm0.03$ & $0.77\pm0.09$  &  $0.47\pm0.02$ & $0.76\pm0.05$  &  $0.54\pm0.02$ & $0.69\pm0.03$ \\
Orthogonal         &  $0.38\pm0.02$ & $0.71\pm0.05$  &  $0.47\pm0.04$ & $0.71\pm0.11$  &  $0.48\pm0.02$ & $0.73\pm0.06$  &  $0.54\pm0.01$ & $0.59\pm0.04$ \\
Reduced major axis &  $0.37\pm0.02$ & $0.75\pm0.05$  &  $0.46\pm0.03$ & $0.77\pm0.09$  &  $0.47\pm0.02$ & $0.76\pm0.05$  &  $0.54\pm0.02$ & $0.67\pm0.04$ \\
Mean OLS           &  $0.37\pm0.02$ & $0.75\pm0.05$  &  $0.45\pm0.03$ & $0.79\pm0.09$  &  $0.47\pm0.02$ & $0.77\pm0.05$  &  $0.54\pm0.02$ & $0.71\pm0.04$ \\
\hline
\end{tabular}
\end{table*}

\begin{table*}
\caption{
Linear regression coefficients for the $K$-band disc size-luminosity relations
of different disc galaxy types: 
$\log h = a - 0.4b(M_{\rm disc}+24)$. 
\label{Tab_KLR}
}
\begin{tabular}{lcc cc cc cc}
\hline
Method & \multicolumn{2}{c}{$T=1\pm1$ (Sa)} & \multicolumn{2}{c}{$T=3\pm1$ (Sb)} & \multicolumn{2}{c}{$T=5\pm1$ (Sc)} & \multicolumn{2}{c}{$T\geq6$ ($\geq$Scd)} \\
                   &       $a$      &      $b$       &         $a$    &       $b$      &         $a$    &      $b$       &       $a$      &     $b$   \\
\hline
OLS(Y$\mid $X)     &  $0.51\pm0.03$  &  $0.53\pm0.06$  &  $0.55\pm0.02$  &  $0.48\pm0.04$  &  $0.58\pm0.02$  &  $0.34\pm0.03$  &  $0.61\pm0.03$  &  $0.21\pm0.05$ \\
OLS(X$\mid $Y)     &  $0.51\pm0.04$  &  $1.15\pm0.17$  &  $0.54\pm0.02$  &  $0.92\pm0.07$  &  $0.64\pm0.02$  &  $0.73\pm0.07$  &  $0.87\pm0.08$  &  $0.63\pm0.12$ \\
OLS bisector       &  $0.51\pm0.03$  &  $0.80\pm0.07$  &  $0.54\pm0.02$  &  $0.68\pm0.04$  &  $0.61\pm0.01$  &  $0.52\pm0.03$  &  $0.73\pm0.03$  &  $0.40\pm0.04$ \\
Orthogonal         &  $0.51\pm0.03$  &  $0.70\pm0.09$  &  $0.55\pm0.02$  &  $0.58\pm0.05$  &  $0.58\pm0.02$  &  $0.39\pm0.04$  &  $0.62\pm0.03$  &  $0.23\pm0.05$ \\
Reduced major axis &  $0.51\pm0.03$  &  $0.78\pm0.08$  &  $0.54\pm0.02$  &  $0.67\pm0.04$  &  $0.60\pm0.01$  &  $0.50\pm0.03$  &  $0.70\pm0.03$  &  $0.36\pm0.03$ \\
Mean OLS           &  $0.61\pm0.03$  &  $0.84\pm0.09$  &  $0.54\pm0.02$  &  $0.70\pm0.04$  &  $0.61\pm0.01$  &  $0.53\pm0.04$  &  $0.74\pm0.04$  &  $0.42\pm0.05$ \\
\hline
\end{tabular}
\end{table*}

From Graham \& Driver (2005, their equation~12) one has that the absolute magnitude 
\begin{equation}
M = \langle \mu \rangle_{\rm e} - 2.5\log[2\pi R_{\rm e}^2] -
36.57, 
\end{equation}
with $R_{\rm e}$ in kpc.  Therefore, for a disc with $n=1$, 
\begin{equation}
M_{\rm disc} = (\mu_0 + 1.123) - 2.5\log[2\pi (1.678 h)^2] - 36.57. 
\label{Eq_11}
\end{equation}
From this relation one can 
readily express the disc size-luminosity relations (Tables~\ref{Tab_LR} and
\ref{Tab_KLR}) in terms of size and surface 
brightness.  For example, combining equations~\ref{Eq_LR} and \ref{Eq_11} to
eliminate $M_{\rm disc}$, one has 
\begin{equation}
\mu_0 = \frac{5b-2.5}{b} \log h - M_0 + \frac{2.5a}{b} + 38.57. 
\end{equation}

If the typical central surface brightness $\mu_0 = -2.5 \log I_0$ was constant
for the discs, although we know from Figures~\ref{fig-muer} and
\ref{fig-murky} that it is not, then one would find $h \sim L^{0.5}$.  While a
symmetrical regression of the optical data suggests $h \propto
L^{0.75\pm0.04}$ for the discs of early-type disc galaxies (Table~\ref{Tab_LR}),
the near-infrared data has a slope which is smaller, albeit only at the
$\sim$1.5$\sigma$ level.

A quick visual inspection of Figure~\ref{Fig_LR} reveals that 
fitting to the full data set, i.e.\ both the early- and late-type
disc galaxies, would produce $h \sim L^b$ relations with shallower slopes
(smaller $b$ values) 
than those reported in Table~\ref{Tab_LR}. 
We have not done this because (i) sample selection effects may have introduced
an artificial boundary to the distribution of the late-type disc galaxies and
(ii) from the $\mu_0 - \log h$ diagrams it was already clear that no real relation
exists for the late-type disc galaxy population.  

Before contrasting our results with others, we note that qualitative
differences should not be surprising.  We have derived intrinsic, fully dust
corrected, {\it disc} scalelengths and luminosities from S\'ersic $R^{1/n}$
bulge plus exponential disc decompositions.  We first compare our $K$-band
results with Courteau et al.\ (2007) who presented $K$-band {\it galaxy}
luminosity-size relations for 360 2MASS galaxies binned into different galaxy
types.
%
At least for systems with small bulge-to-disc flux ratios, the distribution of
disc light is roughly approximated by the distribution of the galaxy light.
Indeed, as shown in Graham (2002, his Figure~6), galaxy half light radii are
roughly equal to the half-light radii of the disc component.  

Courteau et al.\ (2007) used an orthogonal regression analysis which 
can be seen in Tables~\ref{Tab_LR} and \ref{Tab_KLR}
to give a shallower relation to our data than the three other symmetrical 
regression analyses which we used.  
%
For our (potentially sample-selection biased) late-type galaxies ($T>5$) we
have the $K$-band relation $h \sim L^{0.23\pm0.05}$ (orthogonal regression), 
which is in fair agreement with Courteau et al.'s relation $L_{\rm gal}^{0.31\pm0.17}$
for Sd galaxies (their Table~3).  For our Sc, Sb and Sa galaxy samples we find
exponents of $0.39\pm0.04$, $0.58\pm0.05$ and $0.70\pm0.09$ respectively.
Such an increase is in qualitative agreement with 
Courteau et al.\ who report values of $0.39\pm0.03$, $0.43\pm0.04$ and $0.57\pm0.04$.

Courteau et al.\ additionally presented a size-luminosity diagram for 1300
disc galaxies with $I$-band imaging, reporting 
$h \sim L^{0.33\pm0.02}$ (Sc galaxies), 
$h \sim L^{0.37\pm0.02}$ (Sb galaxies) and 
$h \sim L^{0.55\pm0.05}$ (Sa galaxies).  
When we fit all of our $I$-band data, i.e.\ late- and early-type disc galaxies, 
we obtain $h \sim L^{0.45\pm0.05}$ (orthogonal regression), 
while our fit to the S0-Sbc galaxies is notably steeper, 
with $h \sim L^{0.73\pm0.06}$ (orthogonal regression). 
Due to our smaller number of galaxies with $I$-band data, we do not attempt a
finer division of galaxy type. 
%

We have also compared our result with the disc galaxy size-luminosity
relations in Shen et al.\ (2003).  In their work they show the $r^{\prime}$
{\it galaxy} luminosity versus the {\it galaxy} half light radius.  At the
luminous end of Shen et al.'s size-luminosity diagram ($M_{r^{\prime}} < -20$
mag), they report that radius varies with $L^{0.53}$.  They performed an
OLS(Y$\mid$X) regression analysis on binned data, making it additionally
difficult to perform an exact comparison with our results.  At magnitudes
brighter than $M_{r^{\prime}} = -20$ mag, their Figure~5 reveals that the
early-type disc galaxies follow the relation $h \sim L^{0.58}$.  At
lower-luminosities they report shallower slopes.

Shen et al.\ (2003) assumed that galaxies with concentration parameters less
than 2.86 are disc galaxies.  When dealing with bright galaxy samples this is
a reasonable approximation which has been illustrated with visually classified
samples.  However, a problem occurs when one expands one's sample to fainter
magnitudes because most dwarf elliptical galaxies have $n<2$ and thus
concentrations less than $\sim 2.86$ (Graham et al.\ 2005, their table~1).
As a result, faintward of $\sim$ $-19$ $r^{\prime}$-mag ($\sim$ $-20$
$z^{\prime}$-mag) in Shen et al.'s Figures~4-7 (and 10), one does not know if
the systems are actually early- or late-type galaxies.  This can however be
resolved by looking at Shen et al.'s Figures~8-9, in which it becomes apparent
that the curved nature of their disc galaxy size-luminosity relation is due to
the mixing of galaxy types at the low-luminosity end.  This implies that (i)
the curved size-luminosity relations for late-type galaxies in Table~1 from
Shen et al.\ are not appropriate and (ii), the linear size-luminosity
relations for early-type galaxies in their Table~1 are also not appropriate at
faint magnitudes.
%
This has prompted us to revisit the size-luminosity diagram for elliptical 
galaxies, to which we are able to add and compare the distribution of bulges. 

\subsubsection{Elliptical galaxies}

\begin{figure}
\includegraphics[angle=270,scale=0.7]{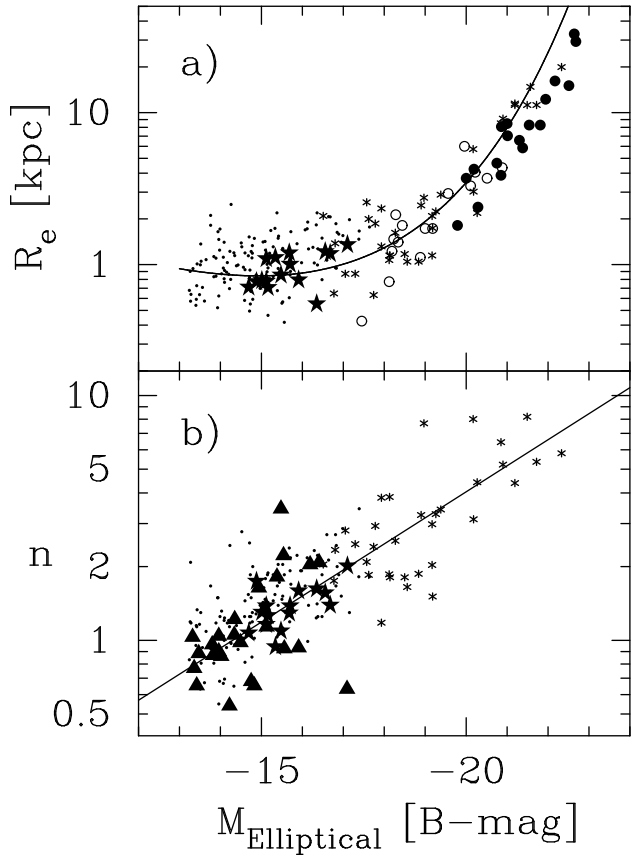}
\caption{ $B$-band. 
Panel a) Luminosity-size diagram for elliptical galaxies.  The data have come
from the compilation in Graham \& Guzm\'an (2003, see their Figure~9).  The
line is not a fit to the data, but a prediction based upon the linear
relations 
between absolute magnitude and central surface brightness (Graham \& Guzm\'an
2003, their Figure~9f) and absolute magnitude and the logarithm of the 
S\'ersic index --- shown here in panel b) for convenience but taken from
Graham \& Guzm\'an (2003, their Figure~10).
A fixed change in S\'ersic index $n$, for example $\delta n = 1$, causes a large
change to the profile when $n$ is small, but only a small change when $n$ is
large.  Because of this, plots of galaxy parameters such as magnitude or
colour will appear curved (flattening at large $n$) rather than linear, if they
are plotted against $n$ rather than $\log n$.
}
\label{Fig_LR_Ell}
\end{figure}

From $-18 < M_B < -13$ mag, the effective radii of dwarf elliptical galaxies
display no real trend with magnitude but instead scatter around a value of
$\sim$1 kpc (Smith Castelli et al.\ 2008; Dabringhausen et al.\ 2008;
Figure~\ref{Fig_LR_Ell}a).  This often over-looked result was previously shown
in Binggeli \& Jerjen (1998) and actually also noted in Shen et al.\ (2003,
their Figures~8 and 9) but the implications of which were not propagated into
all of their diagrams.

The curved distribution of data points in Figure~\ref{Fig_LR_Ell}a can be
predicted (see the line in this figure) from the following linear $M - \mu_0$
and $M - \log n$ (Figure~\ref{Fig_LR_Ell}b) relations for elliptical galaxies.
Using equations~7, 9 and 12 from Graham \& Driver (2005), one has an
expression involving $M, R_{\rm e}, \mu_0$ and $n$ for any S\'ersic profile.
For elliptical galaxies, the $\mu_0$ term in this (not shown) expression 
can be replaced with $M$ via the
empirical relation
\begin{equation}
M_B = \frac{2}{3}\mu_{0,B} -29.5
\end{equation}
from Graham \& Guzm\'an (2003), to give the relation 
\begin{equation}
\log R_e [{\rm kpc}] = 1.137 + 0.217b_n + \frac{M_B}{10} + 0.5\log\left[
  \frac{b_n^{2n}} {n\Gamma (2n){\rm e}^{b_n}} \right], 
\end{equation}
where $b_n \approx 1.9992n-0.3271$, for $0.5 < n < 10$ (Capaccioli 1989). 
The value of $n$ can also be expressed in terms of $M_B$ using the 
empirical relation 
\begin{equation}
n = 10^{ -(14.3+M_B)/9.4}
\end{equation}
from Graham \& Guzm\'an (2003).   This is how the line in
Figure~\ref{Fig_LR_Ell}a was
derived.  It can be seen to match the behaviour of the data rather well, 
which has (only) an approximately linear behaviour ($R_{\rm e}\propto L^{0.9}$ from 
the data in Figure~\ref{Fig_LR_Ell}a) at luminosities brighter 
than $M_B \sim -19$ mag.  
%
%
%
The gradual curvature in the bright end of this $L-R$ relation is
independently supported by the data in Liu et al.\ (2008, their Fig.12), for
which their brightest cluster galaxies (when using 25 $r$-mag/arcsec$^2$
isophotal magnitudes) have $R_{50}\propto L^{0.88}$.  The exponential-like
envelopes of cD galaxies (Seigar et al.\ 2007) does however complicate the
derivation of reliable brightest cluster galaxy parameters and we do not
explore this issue here.

The curved size-luminosity relation for elliptical galaxies, predicted from
two linear relations, is probably not a fundamental relation.  Due to the
continually varying slope of the $L-R$ relation, the mean slope for dwarf and
bright elliptical galaxies will be different.  This is, however, not evidence
for two disjoint species of galaxies.  Similarly, as explained in Graham \&
Guzm\'an (2003), the continually curved size-(effective surface brightness)
relation for elliptical galaxies is not evidence for two distinct types of
galaxy.  Instead, the curved relation is expected from the linear relations
which dwarf and luminous elliptical galaxies are observed to co-define. 

Having derived the curved size-luminosity relation for elliptical galaxies
from two linear relations, we examine why the bright end slope is different to
that reported by Shen et al.\ (2003).  Shen et al.\ used Petrosian quantities
for the luminous elliptical galaxies, however Petrosian magnitudes miss the
outer flux in galaxies, and increasingly so for brighter galaxies.  While the
Petrosian magnitude is therefore fainter than the total magnitude, it is
actually the Petrosian half flux radii $R_{50}$ which is most heavily impacted
in comparison with the effective half light radii $R_{\rm e}$.  This is
quantified in Graham et al.\ (2005).  Looking at Shen et al.'s Figure~4, one
can see that their early-type galaxies display a linear size-luminosity
relation from $-19.75 > M_{r^{\prime}} > -23.75$.  From Fukugita et al.\
(1995) we have $B-r^{\prime} = 1.32$ and so the above magnitude range
corresponds to $-18.43 > M_B > -22.43$.  From Figure~\ref{Fig_LR_Ell}b we see
that this magnitude range corresponds to a range in S\'ersic index from 3 to
7.  We can now use the relation $R_{50} \approx [1 - 6.0\times
10^{-6}(R_{90}/R_{50})^{8.92}]R_{\rm e}$ from Graham et al.\ (2005) to
quantify the difference between the Petrosian half light radii $R_{50}$ (used
by Shen et al.) and the effective half light radii $R_{\rm e}$ (used in
Figure~\ref{Fig_LR_Ell}a).  When $n=7$ ($R_{90}/R_{50}=3.62$), $R_{50} = 0.42
R_{\rm e}$.  When $n=3$ ($R_{90}/R_{50}=3.17$), $R_{50} = 0.82 R_{\rm e}$.
Such a systematic reduction to the radii, as the elliptical galaxy luminosity
increases, results in a reduction (of a factor $\log [4.2/0.82] \approx 0.71$)
to the exponent $b$ in the relation $R\propto L^b$, and likely explains why
Shen et al.\ (2003) find $R_{50} \propto L_{\rm Petrosian}^{0.60}$.
The S\'ersic radii from Blanton et al.\ (2005) --- which Shen et al.\
additionally used --- are also known to systematically underestimate the true
effective half light radii as the S\'ersic index increases (see Figure~9 from
Blanton et al.\ 2005).  Underestimation also occurs when using $R^{1/4}$
models to describe systems with $n>4$ (see Trujillo et al.\ 2001, their
Figure~4) which may explain why Bernardi et al.\ (2003) report that $R \sim
L^{0.63}$.
Therefore, while the above results are not necessarily wrong, it does
highlight that if we are to compare theory (e.g.\ Mayer, Governato \& Kaufmann
2008) and observations, it is important that we measure radii and luminosities
in a physically appropriate and consistent manner.

\subsubsection{Bulges}

\begin{figure}
\includegraphics[angle=270,scale=0.59]{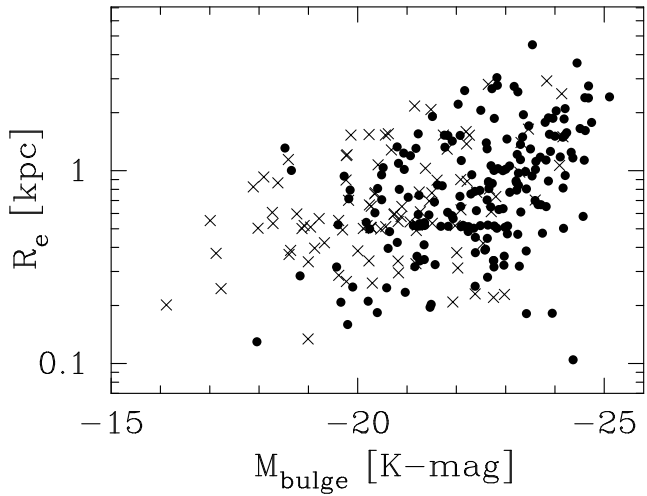}
\caption{ 
$K$-band.  Size-luminosity diagram for bulges. 
Galaxy types Sbc ($T=4$) and earlier are 
denoted by the circles, while later galaxy types are denoted by the crosses.
}
\label{Fig_LR_Bul}
\end{figure}

Finally, with new data in hand, the $K$-band size-luminosity distribution for
the bulges is shown in Figure~\ref{Fig_LR_Bul}.  No real trend is evident and
we refrain from fitting a relation to this scatter diagram.  We can however
explore if the upper boundary in this Figure is real.  
Potentially, just as
low surface brightness disc galaxies avoided detection due to effectively
being hidden by the sky background flux (Disney 1976; Longmore et al.\ 1982;
Davies et al.\ 1988), a population of large, low surface brightness bulges
(which could populate the upper left portion of Figure~\ref{Fig_LR_Bul}) may
be submerged beneath the flux of their associated discs.  
The existence of such bulges is likely to be rare for the simple fact that
galaxies in which no bulge is detected, i.e.\ bulgeless galaxies, are rare.
If a bulge was to exist with, for example, $R_{\rm e} = 3$ kpc and an absolute
$K$-band magnitude of $-20$ mag, then it would have a mean effective surface
brightness of 20.95 $K$-mag arcsec$^{-2}$ (Graham \& Driver 2005, their
eq.12).  If this bulge had a S\'ersic index $n=1$, then it would have a
$K$-band central surface brightness of 19.83 mag arcsec$^{-2}$.  If bright
enough relative to the host disc, this will result in a central protrusion of
the galaxy light profile, above that of the disc profile.  For late-type,
bulgeless disc galaxies (say Sd or later), we can see from Table~\ref{Tab_K}
that their central disc surface brightnesses are not bright enough to hide
such a bulge.  On the other hand, a bulge with $R_{\rm e} = 3$ kpc and an
absolute $K$-band magnitude of $-18$ mag, and thus a central surface
brightness of 21.83 mag arcsec$^{-2}$, could be hidden.  Although we note that
if the ratio $<$$R_{\rm e}/h$$> \approx 0.2-0.25$ holds, it would require the
late-type disc galaxy has a scalelength around 12-15 kpc, something which is
not seen in Figures~\ref{fig-muer} or \ref{fig-murky}.  We therefore
tentatively conclude that the upper boundary in Figure~\ref{Fig_LR_Bul} is
real.

\begin{figure}
\includegraphics[angle=270,scale=0.59]{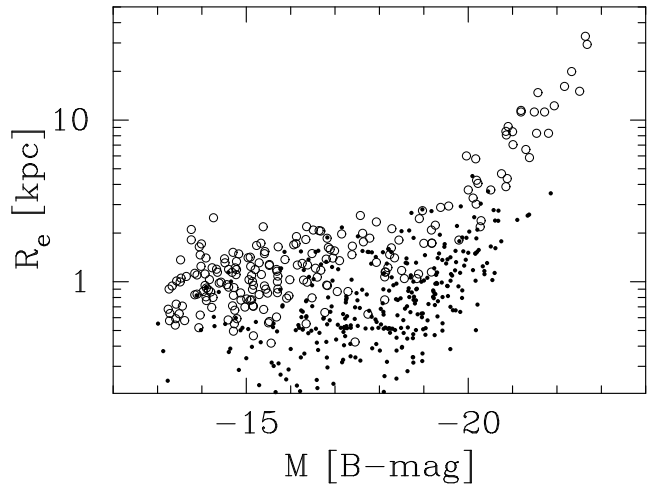}
\caption{ 
Size-luminosity diagram for elliptical galaxies (open circles) and
bulges (dots).  
The $K$-band bulge magnitudes have been 
shifted to the $B$-band using a constant $B-K$ colour of 4.0. 
}
\label{Fig_LR_both}
\end{figure}

In Figure~\ref{Fig_LR_both} we have very crudely adjusted the $K$-band bulge
magnitude to the $B$-band by using $B-K=4$, so as to place the bulges in the
same diagram as elliptical galaxies.  While they roughly overlap with the
dwarf elliptical galaxies, at a given magnitude they scatter to smaller sizes.
This is somewhat reminiscent of their behaviour in the magnitude - S\'ersic
index diagram, where, for a given magnitude, the bulges scatter to smaller
S\'ersic indices than the dwarf elliptical galaxies do (Graham 2001a, his
Fig.14).

\section{Summary} 

Most bright galaxies in the Universe today are spiral galaxies. 
Loveday (1996) has estimated that two-thirds of galaxies brighter than $M_V =
-18$ mag are spiral galaxies and Lintott et al.\ (2008) 
report that the elliptical to spiral ratio is 0.57 for galaxies with 
$M_r < -19$ mag; the actual number of disc (i.e.\ spiral plus lenticular)
galaxies is greater still.  Discs are well known to contain dust which biases
our measurements of their physical parameters and also that of their embedded
bulges. 
Matching observational data from over 10,000 galaxies (Allen et al.\ 2006) 
with radiative transfer 
measurements from sophisticated, three-dimensional models of disc galaxies
with dust 
(Popescu et al.\ 2000; Tuffs et al.\ 2004), Driver et al.\ (2007) has recently
provided a highly useful calibration of these models.  From this, equations to correct
the observed scalelength and surface brightness for inclination and dust are
derived and presented here (equations~\ref{Eq_mu} and \ref{Eq_h}).  
In the optical bands, dust can bias the measured disc scalelength and surface
brightness by 40 per cent and $\sim$0.5 mag arcsec$^{-2}$, respectively, while
the disc magnitude can be in error by up to 1 magnitude.  Corrections for dust
are therefore quite significant.  We remark that programs to measure the
evolution of disc galaxy parameters may need to additionally consider
the evolution of their dust.

These attenuation-inclination corrections enabled us to provide a modern, 
quantitative measure of the intrinsic physical properties of disc galaxies in
the local ($z <$ 0.03--0.04) Universe.  We have done this by converting the
observed properties of disc galaxies into their intrinsic (not simply face-on)
dust free values.  This is desirable because comparison between theory and
observations, which is used to facilitate our understanding of galaxy
formation and evolution, is hampered when one's observational measurements are
compromised by the attenuating effects of dust.

Using a sample of galaxies which have been carefully fitted with S\'ersic
$R^{1/n}$ bulges plus exponential discs, this paper presents intrinsic
structural properties for disc galaxies as a function of galaxy type.  This
work improves upon measurements obtained using $R^{1/4}$ models for 
every bulge, and
provides inclination- {\it and} dust-corrected disc scalelengths, central
surface brightnesses and magnitudes in optical and near-infrared passbands.

In addition, we present inclination- and dust-corrected bulge-to-disc flux
ratios in several passbands, and $K$-band bulge-to-disc size ratios, as a
function of morphological type.  For a given galaxy type, 
the logarithm of the $B$-band bulge-to-disc
flux ratios from Simien \& de Vaucouleurs (1986) are
as much as $\sim$0.5 dex larger than the values obtained here. 
This is predominantly due to the best-fitting S\'ersic $R^{1/n}$ 
bulge models having $n < 4$, while Simien \& de Vaucouleurs force fit an 
$R^{1/4}$ model to every bulge. 
In general, the median dust-corrected bulge-to-total flux (mass) ratios are
less than 1/3 (1/2), which may provide a useful new 
restraint on some simulations which tend to produce larger 
ratios.   Given the observed range of bulge S\'ersic indices from $\sim$0.5 to $\sim$4, 
we advocate the use of Prugniel \& Simien's (1997) density model 
which was developed to match deprojected S\'ersic profiles, rather than only 
matching a deprojected $R^{1/4}$ profile.

Rather than only dividing the galaxies into early- and late-type bins 
based upon their colour and/or concentration index, catalogued Hubble types 
were available for all of our galaxies.  
This has also enabled us to perform our analysis on a sample of disc galaxies,
rather than an unknown blend of early- and late-type galaxies which have
similar concentration indices at low magnitudes.  
However no attempt has been made to measure volume densities given the ad-hoc
galaxy selection criteria.  Although we note that in Driver et al.\ (2007, 2008),
which contains expressions to correct bulge and disc magnitudes in a range of optical
and near-infrared passbands, including the Sloan Digital Sky Survey (SDSS;
York et al.\ 2000) filter set, we have already done this for the luminosity
density of bulges and discs.

We have presented updated (surface brightness)-size diagrams for bulges and
discs and investigated the size-luminosity relations.  We reveal that the
discs of late-type disc galaxies occupy a broad region in these diagrams and
are not well represented by linear relations.  In contrast, the discs of
early-type disc galaxies appear to reside over a narrower territory, defining
the bright envelope to these distributions.  We have additionally presented
the size-luminosity relation for bulges and elliptical galaxies, emphasizing
that a linear relation is inappropriate.  The bulges of disc galaxies are
shown to overlap with, but additionally scatter to smaller sizes than, the
distribution of similarly luminous elliptical galaxies.

To conclude, while the amount of dust in galaxy discs is roughly one thousand
times less than the stellar mass, it can have a large influence on the
observed properties of galaxies.  In this study we have corrected for the
influence of dust in order to obtain the intrinsic stellar properties of
galaxies.  Such characteristic physical properties, and indeed their bivariate
distributions and possible scaling relations, are not only interesting in
their own right but provide valuable $z=0$ constraints for models of galaxy
evolution.

\section{acknowledgments}

We thank Cristina Popescu and Richard Tuffs for kindly proof-reading 
this manuscript and providing helpful feedback.
We also thank Ileana Vass for kindly faxing us select pages from the RC3. 
We are grateful to both the Australian National University summer school 
program which initiated this project in late 2005, and to the 
Swinburne University of Technology's Researcher Development Scheme
which enabled completion of this project in 2007. 
This research has made use of the NASA/IPAC Extragalactic Database (NED),


\label{lastpage}

\begin{thebibliography}{99999}

\bibitem{Aba3a}Abadi, M.G., Navarro, J.F., Steinmetz, M., Eke, V.R.\ 2003a, ApJ, 591, 499
\bibitem{Aba3b}Abadi, M.G., Navarro, J.F., Steinmetz, M., Eke, V.R.\ 2003b, ApJ, 597, 21
\bibitem{Allen}Allen, P.~D., Driver, S.~P., Graham, A.~W., Cameron, E., Liske, J., \& de Propris, R.\ 2006, MNRAS, 371, 2
\bibitem{Almei}Almeida, C., Baugh, C.~M., Wake, D.~A., Lacey, C.~G., Benson, A.~J., Bower, R.~G., \& Pimbblet, K.\ 2008, MNRAS, 466
\bibitem{And01}Andersen, D.~R., Bershady, M.~A., Sparke, L.~S., Gallagher, J.~S., III, \& Wilcots, E.~M.\ 2001, ApJ, 551, L131
\bibitem{APB95} Andredakis, Y.C., Peletier, R.F., \& Balcells, M.\ 1995, MNRAS, 275, 874
\bibitem{AaS94}Andredakis, Y.C., Sanders, R.H., 1994, MNRAS, 267, 283
\bibitem{ABM06}Arag{\'o}n-Salamanca, A., Bedregal, A.~G., \& Merrifield, M.~R.\ 2006, A\&A, 458, 101 
\bibitem{Athan}Athanassoula, E.\ 2008, IAU Symposium 245 ``Galactic bulges'', M. Bureau et al.\ eds (arXiv:0802.0151)
\bibitem{BaH08}Bailin, J., Harris, W.E.\ 2008, ApJ, in press (arXiv:0803.1274)
\bibitem{Bet03}Balcells, M., Graham, A.~W., Dom{\'{\i}}nguez-Palmero, L., \& Peletier, R.~F.\ 2003, ApJ, 582, L79 
\bibitem{BGP07}Balcells, M., Graham, A.~W., \& Peletier, R.~F., 2007, ApJ, in press (astro-ph/0404381)
\bibitem{BaS03}Barnes, E.I., \& Sellwood, J.A.\ 2003, AJ, 125, 1164
\bibitem{Bar07}Barway, S., Kembhavi, A., Wadadekar, Y., Ravikumar, C.D., Mayya, Y.D., 2007, ApJL, 661, L37
\bibitem{BWY02}Bekki, K., Couch, J., \& Yasuhiro, S.\ 2002, ApJ, 577, 651
\bibitem{Bern3}Bernardi, M., et al.\ 2003, AJ, 125, 1849 
\bibitem{BaJ98}Binggeli, B., Jerjen, H.\ 1998, A\&A, 333, 17
\bibitem{BaM98}Binney, J., \& Merrifield, M.\ 1998, Galactic astronomy / James Binney and Michael Merrifield.~ Princeton, NJ : Princeton University Press, 1998.~ (Princeton series in astrophysics)
\bibitem{Bet02}Blakeslee, J.P., et al., 2002, MNRAS, 330, 443
\bibitem{Blant}Blanton, M.~R., et al.\ 2005, AJ, 129, 2562 
\bibitem{Bois4}Boissier S., Boselli A., Buat V., Donas J., Milliard B., 2004, A\&A, 424, 465
\bibitem{Bor81}Boroson, T.\ 1981, ApJS, 46, 177 
\bibitem{Bow06}Bower, R.G., Benson, A.J., Malbon, R., Helly, J.C., Frenk, C.S., Baugh, C.M., Cole, S., \& Lacey, C.G.\ 2006, MNRAS, 370, 645
\bibitem{Bourn}Bournaud, F., Combes, F., Jog, C.J., \& Puerari, I.\ 2005, A\&A, 438, 507
\bibitem{Brown}Brown, R.J.N., et al.\ 2003, MNRAS, 341, 747
\bibitem{BaC03}Bruzual, G., \& Charlot, S.\ 2003, MNRAS, 344, 1000
\bibitem{CPM08}Calura, F., Pipino, A., Matteucci, F.\ 2008, in XIXemes Rencontres de Blois (arXiv:0801.2551)
\bibitem{CCD93}Caon, N., Capaccioli, M., \& D'Onofrio, M.\ 1993, MNRAS, 265, 1013
\bibitem{Cap89}Capaccioli, M.\ 1989, in The World of Galaxies, ed.\ H.G.\ Corwin, L.\ Bottinelli (Berlin: Springer-Verlag), 208
\bibitem{CCM89}Cardelli, J.~A., Clayton, G.~C., \& Mathis, J.~S.\ 1989, ApJ, 345, 245
\bibitem{Car07}Carollo, C.~M., Scarlata, C., Stiavelli, M., Wyse, R.~F.~G., \& Mayer, L.\ 2007, ApJ, 658, 960 
\bibitem{CaG03}Castro-Rodr{\'{\i}}guez, N. \& Garz\'on, F. 2003, A\&A, 411, 55 
\bibitem{Cio91}Ciotti, L., Pellegrini, S., Renzini, A., \& D'Ercole, A.\ 1991, ApJ, 376, 380 
\bibitem{Cole0}Cole, S., Lacey, C.G., Baugh, C.M., Frenk, C.S.\ 2000, MNRAS, 319, 168
\bibitem{Court}Courteau, S., Dutton, A.~A., van den Bosch, F.~C., MacArthur, L.~A., Dekel, A., McIntosh, D.~H., \& Dale, D.~A.\ 2007, ApJ, 671, 203 
\bibitem{CJB96}Courteau, S., de Jong, R.~S., \& Broeils, A.~H.\ 1996, ApJ, 457, L73 
\bibitem{Cox00}Cox D.P., 2000, Allen's Astrophysical quantities, New York: AIP Press; Springer, p.341
\bibitem{Croft}Croft, R.A.C., Di Matteo, T., Springel, V., Hernquist, L.\ 2008, MNRAS, submitted (arXiv:0803.4003)
\bibitem{Crot6}Croton, D.J., et al.\ 2006, MNRAS, 365, 11
\bibitem{DHK08}Dabringhausen, J., Hilker, M., \& Kroupa, P.\ 2008, MNRAS, in press (arXiv:0802.0703)
\bibitem{Dav88}Davies, J.I., Phillipps, S., \& Disney, M.J.\ 1988, MNRAS, 231, 69p
\bibitem{dJ96b}de Jong, R.S.\ 1996b, A\&A, 313, 45 
\bibitem{dJ96a}de Jong, R.S.\ 1996a, A\&AS, 118, 557
\bibitem{devdK}de Jong, R.S., \& van der Kruit, P.C., 1994, A\&AS, 106, 451
\bibitem{deLu6}de Lucia, G., Springel, V., White, S.D.M., Croton, D., Kauffmann, G.\ 2006, MNRAS, 366,499
\bibitem{deV48}de Vaucouleurs, G., 1948, de Vaucouleurs, G., 1948, Annales d'Astrophysique, 11, 247 
\bibitem{deV57}de Vaucouleurs, G., 1957, AJ, 62, 69
\bibitem{deV58}de Vaucouleurs, G., 1958, ApJ, 128, 465
\bibitem{RC391}de Vaucouleurs, G., de Vaucouleurs, A., Corwin, H.G., Jr., Buta, R.J., Paturel, G., \& Fouque, P.\ 1991, Volume 1-3, XII, 2069 p.7, Springer-Verlag Berlin Heidelberg New York
\bibitem{deVaP}de Vaucouleurs, G., \& Pence, W.D., 1978, AJ, 83, 1163
\bibitem{Dis76}Disney, M.J.\ 1976, Nature, 263, 573
\bibitem{Dis98}Disney, M.J.\ 1998, IAU Colloquium 171, ASP Conf.\ Ser.,J.I. Davies, C. Impey, and S. Phillipps, eds., 170, 11
\bibitem{DdR06}Dong, X.Y., \& De Robertis, M.M.\ 2006, AJ, 131, 1236 
\bibitem{DOngi}D'Onghia, E., Burkert, A., Murante, G., \& Khochfar, S.\ 2006, MNRAS, 372, 1525
\bibitem{DOno1}D'Onofrio, M.\ 2001, MNRAS, 326, 1517 
\bibitem{Dri07}Driver, S.P., Popescu, C.C., Tuffs, R.J., Liske, J., Graham, A.W., Allen, P.D., De Propris, R., 2007, MNRAS, in press (arXiv:0704.2140)
\bibitem{Dri08}Driver, S.P., Popescu, C.C., Tuffs, R.J., Graham, A.W., Liske,
  J., Baldry, I.K., 2008, ApJL, in press (arXiv:0803.4164)
\bibitem{EaB85}Ebneter, K., \& Balick, B.\ 1985, AJ, 90, 183 
\bibitem{EDD88}Ebneter, K., Davis, M., \& Djorgovski, S.\ 1988, AJ, 95, 422 
\bibitem{Ein65}Einasto, J.\ 1965, Trudy Inst.\ Astrofiz.\ Alma-Ata, 5, 87
\bibitem{Erw03}Erwin, P., Vega Beltran, J.C., Graham, A.W., Beckman, J.E., 2003, ApJ, 597, 929
\bibitem{Esk02}Eskridge, P.B., et al.\ ApJS, 143, 73
\bibitem{FaB92}Feigelson, E.D., \& Babu, G.J.\ 1992, ApJ, 397, 55
\bibitem{FaF05}Ferrarese, L., \& Ford, H.\ 2005, Space Science Reviews, 116, 523 
\bibitem{Ferra}Ferrari, F., Pastoriza, M.G., Macchetto, F., \& Caon, N.\ 1999, A\&AS, 136, 269 
\bibitem{Fit99}Fitzpatrick, E.L.\ 1999, PASP, 111, 63
\bibitem{FCT08}Foyle, K., Courteau, S., Thacker, R.\ 2008, MNRAS, in press (arXiv:0803.2716)
\bibitem{Fre70}Freeman, K.C., 1970, ApJ, 160, 811
\bibitem{Gad08}Gadotti, D.A.\ 2008, MNRAS, 384, 420
\bibitem{Gio95}Giovanelli, R., Haynes, M.P., Salzer, J.J., Wegner, G., da Costa, L.N., \& Freudling, W.\ 1995, AJ, 110, 1059 
\bibitem{Gov04}Governato, F., et al.\ 2004, ApJ, 607, 688
\bibitem{Gov07}Governato, F., Willman, B., Mayer, L., Brooks, A., Stinson, G., Valenzuela, O., Wadsley, J., \& Quinn, T.\ 2007, MNRAS, 374, 1479
\bibitem{Gra01}Graham, A.W.\ 2001a, AJ, 121, 820 (Addendum, 2003, AJ, 125, 3398)
\bibitem{Gra1b}Graham, A.W.\ 2001b, MNRAS, 326, 543
\bibitem{Gra02}Graham, A.W.\ 2002, MNRAS, 334, 721
\bibitem{Gra03}Graham, A.W.\ 2003, AJ, 125, 3398
\bibitem{Gra07}Graham, A.W.\ 2007, MNRAS, 379, 711
\bibitem{GadeB}Graham, A.W., \& de Blok, W.J.G.\ 2001, ApJ, 556, 177 
\bibitem{GaD05}Graham, A.W., \& Driver, S.P.\ 2005, PASA, 22(2), 118 (astro-ph/0503176)
\bibitem{GaD07}Graham, A.W., \& Driver, S.P.\ 2007, MNRAS, 380, L15
\bibitem{GaD5b}Graham, A.W., Driver, S.P., Petrosian, V., Conselice, C.J., Bershady, M.A., Crawford, S.M., \& Goto, T.\ 2005, AJ, 130, 1535
\bibitem{GaG03}Graham, A.W., \& Guzm\'an, 2003, AJ, 126, 1787
\bibitem{Gra6a}Graham, A.W., Merritt, D., Moore, B., Diemand, J., \& Terzi\'c, B.\ 2006a, AJ, 132, 2701
\bibitem{Gra6b}Graham, A.W., Merritt, D., Moore, B., Diemand, J., \& Terzi\'c, B.\ 2006b, AJ, 132, 2711
\bibitem{GaP99}Graham, A.W., \& Prieto, M., 1999, ApJ, 524, L23
\bibitem{Gros4}Grosb{\o}l, P., Patsis, P.A., \& Pompei, E.\ 2004, A\&A, 423, 849 
\bibitem{Gut92}Guthrie, B.N.G.\ A\&AS, 1992, 93, 255
\bibitem{HaG84}Haynes, M.P., \& Giovanelli, R.\ 1984, AJ, 89, 758
\bibitem{HCS06}Hernandez, X., \& Cervantes-Sodi, B.\ 2006, MNRAS, 368, 351 
\bibitem{HZA07}Hernandez-Toledo, H., Zendejas-Dominguez, J., \& Avila-Reese, V.\ 2007, AJ, in press (arXiv:0705.2041)
\bibitem{Her90}Hernquist, L.\ 1990, ApJ, 356, 359 
\bibitem{Hol46}Holmberg, E.\ 1946, Medd.\ Lund.\ Astron.\ Obs.\ Ser.\ II.\ No.\ 117
\bibitem{Hub26}Hubble, E.\ 1926, ApJ, 64, 321
\bibitem{IDC97} Iodice, E., D'Onofrio, M., Capaccioli, M.\ 1997, ASP Conf.\ Ser., 116, 841
\bibitem{IDC99} Iodice, E., D'Onofrio, M., Capaccioli, M.\ 1999, ASP Conf.\ Ser., 176, 402
\bibitem{JBF00}Jerjen, H., Binggeli, B., \& Freeman, K.C.\ 2000, AJ, 119, 593
\bibitem{KdJP6}Kassin, S.A., de Jong, R.S., Pogge, R.W.\ 2006, ApJS, 162, 80
\bibitem{KWO86}Kodaira, K., Watanabe, M., \& Okamura, S.\ 1986, ApJS, 62, 703 
\bibitem{KaK04}Kormendy, J., \& Kennicutt, R.C., Jr.\ 2004, ARA\&A, 42, 603 
\bibitem{Kor08}Kormendy, J.\ 2008, in Formation and Evolution of Galaxy Bulges, Proceedings IAU Symposium No. 245, 2007, M. Bureau et al. eds., in press (arXiv:0708.2104)
\bibitem{Kor98}Kornreich, D.A., Haynes, M.P., \& Lovelace, R.V.E.\ 1998, AJ, 116, 2154
\bibitem{Ken85}Kent, S., 1985, ApJS, 59, 115
\bibitem{KDF91}Kent, S.M., Dame, T., \& Fazio, G., 1991, ApJ, 378, 131
\bibitem{KWK00}Khosroshahi, H.G., Wadadekar, Y., \& Kembhavi, A.\ 2000, ApJ, 533, 162
\bibitem{Knap3}Knapen, J.H., de Jong, R.S., Stedman, S., \& Bramich, D.M.\ 2003, MNRAS, 344, 527 
\bibitem{Lav95}Lahav, O., Naim, A., Buta, R.J., Corwin, H.G., de Vaucouleurs, G., et al.\ 1995, Science, 267, 859
\bibitem{LSB05}Laurikainen, E., Salo, H., \& Buta, R., 2005, MNRAS, 362, 1319 
\bibitem{Laur7}Laurikainen, E., Salo, H., Buta, R., \& Knapen, J.\ 2007, MNRAS, 381, 401
\bibitem{Laur6}Laurikainen, E., Salo, H., Buta, R., Knapen, J., Speltincx, T., \& Block, D., 2006, AJ, 132, 2634
\bibitem{Leeuw}Leeuw, L.L., Davidson, J., Dowell, C.D., Matthews, H.E.\ 2008, ApJ, 677, 249
\bibitem{Lzoo8}Lintott, C.J., et al.\ 2008, MNRAS, submitted (arXiv:0804.4483)
\bibitem{Liu08}Liu, F.S., Xia, X.Y., Mao, S., Wu, H., Deng, Z.G.\ 2008, MNRAS, 385, 23
\bibitem{Let82}Longmore, A.J., Hawarden, T.G., Goss, W.M., Mebold, U., \& Webster, B.L.\ 1982, MNRAS, 200, 325
\bibitem{Lov96}Loveday, J., 1996, MNRAS, 278, 1025
\bibitem{MCH03} MacArthur, L.A., Courteau, S., \& Holtzman, J.A., 2003, ApJ, 582, 689 
\bibitem{Mall8}Maller, A.H., Berlind, A.A., Blanton, M.R., Hogg, D.W.\ 2008, ApJ, submitted (arXiv:0801.3286)
\bibitem{Mar01}M\'arquez, I., Lima Neto, G.B., Capelato, H., Durret, F., Lanzoni, B., \& Gerbal, D.\ 2001, A\&A, 379, 767
\bibitem{MGH03}Masters K. L., Giovanelli R., Haynes M. P., 2003, AJ, 126, 158
\bibitem{Mat02}Mathis, H., Lemson, G., Springel, V., Kauffmann, G., White, S.D.M., Eldar, A., \& Dekel, A.\ 2002, MNRAS, 333, 739 
\bibitem{MGK08}Mayer, L., Governato, F, Kaufmann, T.\ 2008, Advanced Science Letters (arXiv:0801.3845)
\bibitem{MaO84}Meisels, A., \& Ostriker, J.~P.\ 1984, AJ, 89, 1451 
\bibitem{Men08}M\'endez-Abreu, J., Aguerri, J.A.L., Corsini, E.M., Simonneau, E.\ 2008, A\&A, submitted (arXiv:0710.5466) 
\bibitem{Merr6}Merritt, D., Graham, A.W., Moore, B., Diemand, J., \& Terzi{\'c}, B.\ 2006, AJ, 132, 2685
\bibitem{Mo998}Mo, H.J., Mao, S., White, S.D.M.\ 1998, MNRAS, 295, 319
\bibitem{Molly}M{\"o}llenhoff, C.\ 2004, A\&A, 415, 63 
\bibitem{MaH01}M\"ollenhoff, C., \& Heidt, J.\ 2001, A\&A, 368, 16
\bibitem{MPT06}M\"ollenhoff, C., Popescu, C.C., \& Tuffs, R.J.\ 2006, A\&A, 456, 941
\bibitem{NaB91}Navarro, J.F., \& Benz, W.\ 1991, ApJ, 380, 320
\bibitem{NaW91}Navarro, J.F., \& White, W.\ 1994, MNRAS, 267, 401
\bibitem{NLC06}Nipoti, C., Londrillo, P., \& Ciotti, L.\ 2006, MNRAS, 370, 681 
\bibitem{Oka05}Okamoto, T., Eke, V.~R., Frenk, C.~S., \& Jenkins, A.\ 2005, MNRAS, 363, 1299
\bibitem{Ost77}Ostriker, J.~P.\ 1977, Proceedings of the National Academy of Science, 74, 1767 
\bibitem{PaS08}Padilla, N.D., Strauss, M.A.\ 2008, MNRAS, submitted (arXiv:0802.0877)
\bibitem{Pat40}Patterson, F.S., 1940, Harvard College Observatory Bulletin, 914, 9
\bibitem{PaB96}Peletier, R.F., \& Balcells, M.\ 1996, AJ, 111, 2238
\bibitem{PaW92}Peletier, R.F. \& Willner, S.P.\ 1992, AJ, 103, 1761
\bibitem{Pop00}Popescu, C.C., Misiriotis, A., Kylafis, N.D., Tuffs, R.J., \& Fischera, J.\ 2000, A\&A, 362, 138 
\bibitem{Pop05}Popescu, C.C. et al.\ 2005, ApJ, 619, L75
\bibitem{PaT07}Popescu, C.C., \& Tuffs, R.J.\ 2007, in Proceedings of the
  lectures given at the Les Houche Winter School ``Astronomy in the
  submillimeter and far infrared domains with the Herschel Space Observatory''
  (arXiv:0709.2310)
\bibitem{Num92}Press, W.H., Teukolsky, S.A., Vetterling, W.T., \& Flannery, B.P., 1992, Numerical recipes (2nd ed.; Cambridge: Cambridge Univ. Press)
\bibitem{Pet01}Prieto, M., Aguerri, J.A.L., Varela, A.M., \& Mu\~noz--Tu\~non, C.\ 2001, A\&A, 367, 405
\bibitem{PaS97}Prugniel, P., \& Simien, F.\ 1997, A\&A, 321, 111
\bibitem{SaT03}Scannapieco, C., \& Tissera, P.B.\ 2003, MNRAS, 338, 880
\bibitem{Reese}Reese, A.S., Williams, T.B., Sellwood, J.A., Barnes, E.I., \& Powell, B.A.\ 2007, AJ, 133, 2846
\bibitem{Rei08}Reichard, T.A., Heckman, T.M., Rudnick, G., Brinchmann, J., Kauffmann, G.\ 2008, ApJ, 677, 186
\bibitem{Rest1}Rest, A., et al.\ 2001, AJ, 121, 2431
\bibitem{RaZ95}Rix, H.-W., \& Zaritsky, D.\ 1995, ApJ, 447, 82
\bibitem{SaG85}Sadler, E.~M., \& Gerhard, O.~E.\ 1985, MNRAS, 214, 177
\bibitem{San61}Sandage, A. 1961, The Hubble Atlas of Galaxies (Washington: Carnegie Inst. Washington)
\bibitem{Sat08}Satyapal, S., Vega, D., Dudik, R.~P., Abel, N.~P., \& Heckman, T.\ 2008, ApJ, 677, 906
\bibitem{Sat07}Satyapal, S., Vega, D., Heckman, T., O'Halloran, B., \& Dudik, R.\ 2007, ApJ, 663, L9
\bibitem{Scan8}Scannapieco, C., Tissera, P.~B., White, S.~D.~M., \& Springel, V.\ 2008, MNRAS, submitted (arXiv:0804.3795)
\bibitem{SFD98}Schlegel, D.~J., Finkbeiner, D.~P., \& Davis, M.\ 1998, ApJ, 500, 525 
\bibitem{SGJ07}Seigar, M.S., Graham, A.W., Jerjen, H.\ 2007, MNRAS, 378, 1575
\bibitem{SaJ98}Seigar, M.S., \& James, P.A.\ 1998, MNRAS, 299, 672
\bibitem{Ser63}S\'ersic, J.-L.\ 1963, Boletin de la Asociacion Argentina de Astronomia, vol.6, p.41
\bibitem{Ser68}S\'ersic, J.-L.\ 1968, Atlas de Galaxias Australes (Cordoba: Observatorio Astronomico)
\bibitem{Sha07}Shao, Z., Xiao, Q., Shen, S., Mo, H.~J., Xia, X., \& Deng, Z.\ 2007, ApJ, 659, 1159
\bibitem{Shen3}Shen, S., Mo, H.~J., White, S.~D.~M., Blanton, M.~R., Kauffmann, G., Voges, W., Brinkmann, J., \& Csabai, I.\ 2003, MNRAS, 343, 978 
\bibitem{Shi08}Shields, J.C., Walcher, C.J., Boeker, T., Ho, L.C, Rix, H.-W., van der Marel, R.P.\ 2008, ApJ, in press (arXiv:0804.4024)
\bibitem{SdeV6}Simien, F., \& de Vaucouleurs, G.\ 1986, ApJ, 302, 564 
\bibitem{Smith}Smith Castelli, A.V., Bassino, L.P., Richtler, T., Cellone, S.A., Aruta, C., Infante, L.\ 2008, MNRAS, in press (arXiv:0803.1630)
\bibitem{SMH5a}Springel, V., Di Matteo, T., \& Hernquist, L.\ 2005a, MNRAS, 361, 776
\bibitem{SMH5b}Springel, V., Di Matteo, T., \& Hernquist, L.\ 2005b, ApJ, 620, L79
\bibitem{Spr01}Springel, V., White, S.D.M., Tormen, G., \& Kauffmann, G.\ 2001, MNRAS, 328, 726
\bibitem{SaH03}Springel, V., \& Hernquist, L.\ 2003, MNRAS, 339, 289 
\bibitem{SaN02}Steinmetz, M., \& Navarro, J.~F.\ 2002, New Astronomy, 7, 155 
\bibitem{TBM07}Temi, P., Brighenti, F., \& Mathews, W.G., 2007, ApJ, 660, 1215
\bibitem{TaG05}Terzi{\'c}, B., \& Graham, A.W.\ 2005, MNRAS, 362, 197 
\bibitem{TaG07}Terzi{\'c}, B., \& Sprague, B.J.\ 2007, MNRAS, 377, 855
\bibitem{Ton01}Tonry, J.L., et al., 2001, ApJ, 546, 681
\bibitem{Tru02}Trujillo, I., Asensio Ramos, A., Rubi{\~n}o-Mart{\'{\i}}n, J.~A., Graham, A.~W., Aguerri, J.~A.~L., Cepa, J., \& Guti{\'e}rrez, C.~M.\ 2002, MNRAS, 333, 510 
\bibitem{TGC01}Trujillo, I., Graham, A.W., \& Caon, N.\ 2001, MNRAS, 326, 869
\bibitem{Tuffs}Tuffs, R.~J., Popescu, C.~C., V{\"o}lk, H.~J., Kylafis, N.~D., \& Dopita, M.~A.\ 2004, A\&A, 419, 821 
\bibitem{Tully}Tully, R.~B., Pierce, M.~J., Huang, J.-S., Saunders, W., Verheijen, M.~A.~W., \& Witchalls, P.~L.\ 1998, Aj, 115, 2264
\bibitem{UaR08}Unterborn, C.T., \& Ryden, B.S.\ 2008, ApJ, submitted (arXiv:0801.2400)
\bibitem{Val90}Valentijn, E.A.\ 1990, Nature, 346, 153
\bibitem{vdB01}van den Bosch F.C.\ 2001, MNRAS, 327, 1334 
\bibitem{sLF07}van Leeuwen, F., Feast, M.W., Whitelock, P.A., \& Laney, C.D.\ 2007, MNRAS, in press (arXiv:0705.1592)
\bibitem{WJB08}Weinzirl, T., Jogee, S., Barazza, F.D.\ 2008, (arXiv:0802.3903)
\bibitem{Whi58}Whitford, A.~E.\ 1958, AJ, 63, 201
\bibitem{Xil99}Xilouris E.M., Byun Y.I., Kylafis N.D., Paleologou E.V., Papamastorakis J., 1999, A\&A, 344, 868
\bibitem{York0}York, D.G., et al.\ 2000, AJ, 120, 1579
\bibitem{YaW75}Yoshizawa, M., \& Wakamatsu, K.\ 1975, A\&A, 44, 363 
\bibitem{YaC94}Young, C.~K., \& Currie, M.~J.\ 1994, MNRAS, 268, L11
\end{thebibliography}
\end{document}